\documentclass[journal]{IEEEtran}
\usepackage{multirow}%added by yxw 20161022
\usepackage{diagbox}%added by yxw 20161022
\usepackage{amssymb}
\usepackage{amsmath}
\usepackage{graphicx}
\usepackage{cite}
\usepackage{citesort}
\usepackage{subfigure}
\usepackage{graphicx,epstopdf}
\usepackage{epsfig}	
\usepackage{cite,graphicx,amsmath,amssymb}
\usepackage{comment}

\usepackage{amssymb}
\usepackage{amsmath}
\usepackage{cite}
\usepackage{url}
\usepackage{xcolor}
\usepackage{cite,graphicx,amsmath,amssymb}
\usepackage{subfigure}
\usepackage{citesort}
\usepackage{fancyhdr}
\usepackage{mdwmath}
\usepackage{mdwtab}
\usepackage{caption}
\usepackage{amsthm}
\usepackage{lipsum}

\newtheorem{theorem}{Theorem}

\newtheorem{lemma}{Lemma}

\newtheorem{corollary}{Corollary}

\newtheorem{remark}{Remark}  % added by yxw 20160929
\newtheorem{proposition}{Proposition}

\makeatletter
\def\ScaleIfNeeded{%
\ifdim\Gin@nat@width>\linewidth \linewidth \else \Gin@nat@width
\fi } \makeatother

\begin{document}

\title{A Unified Framework for Non-Orthogonal Multiple Access}

\author{ Xinwei~Yue, Zhijin~Qin,~\IEEEmembership{Member,~IEEE,}
 Yuanwei~Liu~\IEEEmembership{Member,~IEEE,} Shaoli~Kang
 and Yue Chen,~\IEEEmembership{Senior Member,~IEEE}

\thanks{X. Yue is with School of Information and Communication Engineering, Beijing Information Science and Technology University, Beijing 100101, China (email: xinwei.yue@bistu.edu.cn).}
\thanks{Z. Qin is with the School of Computing and Communications, Lancaster University, Lancaster LA1 4YW, U.K. (e-mail: zhijin.qin@lancaster.ac.uk).}
\thanks{Y. Liu and Y. Chen are with School of Electronic Engineering and Computer Science, Queen Mary University of London, London E1 4NS, U.K. (email: \{yuanwei.liu and yue.chen\}@qmul.ac.uk).}
\thanks{S. Kang is with School of Electronic and Information Engineering, Beihang University, Beijing 100191,
China and also with State Key Laboratory of Wireless Mobile Communications, China Academy of Telecommunications Technology(CATT), Beijing 100094, China (email: kangshaoli@catt.cn). Part of this work will be presented in IEEE ICC 2018 \cite{Yue2018framework}.}
}

\maketitle

\begin{abstract}
This paper proposes a unified framework of non-orthogonal multiple access (NOMA) networks. Stochastic geometry is employed to model the locations of spatially NOMA users.
The proposed unified NOMA framework is capable of being applied to both code-domain NOMA (CD-NOMA) and power-domain NOMA (PD-NOMA).
Since the detection of NOMA users mainly depend on efficient successive interference cancellation (SIC) schemes, both imperfect SIC (ipSIC) and perfect SIC (pSIC) are taken into account. To characterize the performance of the proposed unified NOMA framework, the exact and asymptotic expressions of outage probabilities as well as delay-limited throughput for CD/PD-NOMA with ipSIC/pSIC are derived. In order to obtain more insights, the diversity analysis of a pair of NOMA users (i.e., the $n$-th user and $m$-th user) are provided. Our analytical results reveal that: i) The diversity orders of the $m$-th and $n$-th user with pSIC for CD-NOMA are $mK$ and $nK$, respectively; ii) Due to the influence of residual interference (RI), the $n$-th user with ipSIC obtains a zero diversity order; and iii) The diversity order is determined by the user who has the poorer channel conditions out of the pair. Finally, Monte Carlo simulations are presented to verify the analytical results: i) When the number of subcarriers becomes lager, the NOMA users are capable of achieving more steep slope in terms of outage probability; and ii) The outage behavior of CD-NOMA is superior to that of PD-NOMA.
\end{abstract}

\begin{keywords}
A unified framework, imperfect/perfect successive interference cancellation, non-orthogonal multiple access, stochastic geometry
\end{keywords}
%\textcolor[rgb]{0.00,0.00,1.00}{}

\section{Introduction}
With the rapid increase of requirement for the Internet-enabled smart devices, applications and services, the fifth generation (5G) mobile communication networks have sparked a great deal of attention in both academia and industry. The application of new radio scenarios \cite{5Gscenarios}, i.e., ultra-reliable and low latency communications, massive machine type communications and enhanced mobile broadband, aims to satisfy the different requirements for 5G networks \cite{Qin7366613,Qin8350399}. In particular,
the design of novel multiple access (MA) is desired to enhance spectrum efficiency and massive connectivity.
Non-orthogonal multiple access (NOMA) \cite{Ding2017Mag} has been identified as one of the key technologies in 3GPP Long Term Evolution, which has been standard application in downlink multiuser superposition transmission scenarios \cite{MUST1}.
The primary feature of NOMA is its capability of achieving the higher spectrum efficiency, in which multiple users' signals are linearly superposed over different power levels by using the superposition coding scheme \cite{Tse2005fundamentals}, and then transmitted in the same time/frequence resource element (RE). To get the desired signal,
multi-user detection algorithm \cite{Cover1991Elements,Multiuser1998} (e.g., successive interference cancellation (SIC) or message passing algorithm) is carried out at the receiver.

Up to now, NOMA techniques have been investigated extensively. Based on spreading signature of MA, NOMA schemes can be divided into two categories: power-domain NOMA (PD-NOMA) and code-power NOMA (CD-NOMA)\footnote{The superposition of signals for multiple users can be mapped to single subcarrier or multiple subcarriers. Driven by this,
NOMA can also be classified as single carrier NOMA (SC-NOMA) and multi-carrier NOMA (MC-NOMA). More specifically, SC-NOMA and MC-NOMA are equivalent to PD-NOMA and CD-NOMA, respectively.}.
More particularly, the point-to-point PD-NOMA has been surveyed in detail in \cite{Ding2014performance,Pairing7273963,Yang7361990,Ding7906532Multicast}. Two evaluation metrics of PD-NOMA networks including outage probability and ergodic rate have been proposed in \cite{Ding2014performance}, where the outage behaviors of users and ergodic rate have been discussed by applying stochastic geometry. Furthermore, the impact of user pair with fixed power allocation for PD-NOMA has been characterized in terms of outage probability in \cite{Pairing7273963}. It has been shown that when the selected user pairing have more disparate channel conditions, PD-NOMA is capable of providing more performance gain.
From a practical perspective, the authors in \cite{Yang7361990} studied the performance of PD-NOMA for the two-user case with imperfect channel state information, where the closed-form and approximate expressions of outage probability and ergodic sum rate were derived, respectively. On the condition that the NOMA users have similar channel conditions, the authors of \cite{Ding7906532Multicast} proposed a PD-NOMA based multicast-unicast scheme and verified that the spectral efficiency of PD-NOMA based multicast-unicast scheme is higher than that of orthogonal multiple access (OMA) based one.
To evaluate the performance of uplink PD-NOMA, in \cite{Tabassum7913656}, the coverage probability performance of the NOMA users was discussed in large scale cellular for uplink PD-NOMA by invoking poisson cluster processes, where both imperfect SIC (ipSIC) and perfect SIC (pSIC) were taken into considered. By applying the concept of NOMA to cooperative communications, cooperative NOMA was first introduced in \cite{Ding2014Cooperative}, where the nearby users with better channel conditions were regarded as decode-and-forward relay to deliver the signals for the distant users.
To further improve spectrum efficiency, the authors of \cite{Yue8026173} studied the outage behavior and ergodic rate of PD-NOMA, where user relaying can switch between full-duplex mode and half-duplex mode based on application requirements.

As adopted by many 5G MA concepts, CD-NOMA mainly include sparse code multiple access (SCMA), pattern division multiple access (PDMA), multi-user sharing access (MUSA), interleave division multiple access (IDMA), etc. Actually, CD-NOMA is viewed as a special extension of PD-NOMA, in which the data streams of multiple users are directly mapped into multiple REs (or $K$ subcarriers) through the sparse matrix/codebook or low density spread sequence.
More specifically, in \cite{Nikopour2013Sparse}, the modulation symbols of NOMA users are directly mapped into sparse codebook by invoking multidimensional constellation, where a sub-optimal design approach was proposed to design the sparse codebook of SCMA. Considering user pair and power sharing, the system throughput of heavily loaded networks has been improved in \cite{Nikopour7037423SCMA} by adopting SCMA for donwlink transmission scenarios. To perform the green analysis of SCMA scheme, the authors in \cite{Zhang7037563} have analyzed the energy efficiency and outage behavior by proposing the unified framework in fading channels. With the goal of maximizing the ergodic sum rate, an optimal sparse matrix of SCMA system has been designed in \cite{Ding7582475SCMA}. Moreover, the performance of uplink SCMA system has been characterized in terms of average symbol error rate with randomly deployed users in \cite{Bao7983400}.

For another special case, the thought of PDMA was first proposed in \cite{ChenPattern7526461}, where the joint design of sparse matrix and SIC based detector has been considered at the transmitting end and receiving end, respectively. From the perspective of link level and system level, the evaluated results confirmed that PDMA is capable of achieving the enhanced spectrum efficiency over OMA. In the case of given sparse matrix, a novel link estimation scheme for uplink PDMA systems was proposed in \cite{Ren7990128} based on interference cancellation receiver. It was shown that the proposed estimation scheme can achieve accurate performance compared to conventional method. With the aid of pSIC, the authors of \cite{Tang7880556} studied the outage behavior of cooperative uplink PDMA systems by employing one fixed dimension of sparse matrix. As the further special cases \cite{Yuan2016Multi,Ping1618943}, in \cite{Yuan2016Multi}, the data symbols of each user for MUSA systems are spread to a set of complex spreading sequences and then superposed at transmitter. The design of low-correlation spreading sequence is to deal with the higher overloading of users and to carry out SIC expediently at receiver. Exploiting the low-rate coded sequence, the bit error rate of IDMA systems based on semi-analytical technique has been discussed in  \cite{Ping1618943}. Furthermore, the performance of cooperative IDMA systems is characterized in terms of bit error probability in \cite{Bilim7493674}. Very recently, the progresses of CD/PD-NOMA for 5G networks have been surveyed in \cite{Islam7676258,Liu8114722,Cai2017Modulation}, which have summarized potentials and challenges from the perspective of implementation.
% 介绍当前的有关unified NOMA的文章

\subsection{Motivations and Contributions}
While the above-mentioned research contributions have laid a solid foundation for a good understanding of PD-NOMA and CD-NOMA techniques, a unified framework for NOMA networks is far from being well understood. In \cite{Ding2014performance}, it is demonstrated that the diversity order of the sorted NOMA user, i.e., the $m$-th user is $m$, which is directly combined with the users' channel conditions. However, only the performance of PD-NOMA has been discussed. In \cite{Qin7880556Yue}, the authors have proposed user association and resource allocation schemes for the unified NOMA enabled heterogeneous ultra-dense networks.
Moreover, the above contributions for NOMA networks have comprehensively concentrated on the assumption of pSIC. In practice, the assumption of pSIC might not be valid at receiver, since there still exist several potential implementation issues by using SIC, i.e., error propagation and complexity scaling. Hence it is significant to examine the detrimental impacts of ipSIC for the unified NOMA framework. To the best of our knowledge, there is no existing work investigating the unified NOMA network performance, which motivates us to develop this treatise. In addition, new connection outage probability (COP) is defined as an evaluation metric for the unified NOMA framework. The essential contributions of our paper are summarized as follows:
\begin{enumerate}
  \item
   We derive the exact expressions of COP for a pair of users, i.e., the $n$-th user and $m$-th user in CD-NOMA networks.
   Based on the analytical results, we also derive the asymptotic COP and obtain the diversity orders. We confirm that the diversity order of the $m$-th user is equal to $mK$. Due to the impact of residual interference (RI) from the imperfect cancellation process, the COP of the $n$-th user with ipSIC for CD-NOMA converges to an error floor in the high signal-to-noise ratio (SNR) region and obtain a zero diversity order.
   \item
   We study the COP of the n-th user with pSIC and derive the corresponding asymptotic COP for CD-NOMA. On the condition of pSIC, the $n$-th user is capable of attaining the diversity order of $nK$. We confirm that the outage performance of the $n$-th user with pSIC is superior to OMA, while the outage performance of the $m$-th user is inferior to OMA. It is shown that when multiple users are served simultaneously, NOMA is capable of providing better fairness.
   \item
   We investigate the outage behaviors of the special case PD-NOMA with ipSIC/pSIC for CD-NOMA ($K=1$). To provide valuable insights, we derive both exact and asymptotic COP of a pair of users for PD-NOMA. We observe that the diversity orders of the $n$-th user with ipSIC/pSIC are equal to $n$ and zero, respectively. The $m$-th user of PD-NOMA obtains the diversity order of $m$.
    \item
   For the selected user pairing in CD/PD-NOMA networks, we observe that the diversity order is determined by the user who has the poorer channel conditions out of the pair. We discuss the system throughput of CD/PD-NOMA with ipSIC/pSIC in delay-limited transmission mode. When frequency dependent factor $\eta = 1$, we observe that the outage performance of the $n$-th user with ipSIC is superior to OMA in the low SNR region.
\end{enumerate}

\subsection{Organization and Notation}
The remainder of this paper is organized as follows. In Section \ref{Network Model}, a unified NOMA framework is presented
in the wireless networks, where users are ordered randomly based on their channel conditions. In Section \ref{Connection Outage},
the exact expressions of outage probability and delay-limited throughput for a pair of NOMA users are derived. Section \ref{Numerical Results} provides the numerical results to verify the derived analytical results and then Section \ref{Conclusion} concludes our paper. The proofs of mathematics are collected in the Appendix.

The main notations of this paper are shown as follows:
$\mathbb{E}\{\cdot\}$ denotes expectation operation; ${f_X}\left(  \cdot  \right)$ and ${F_X}\left(  \cdot  \right)$ denote the probability density function (PDF) and cumulative distribution function (CDF) of a random variable $X$, respectively; The superscripts ${\left(  \cdot  \right)^T}$ and ${\left(  \cdot  \right)^*}$ stand for transpose and conjugate-transpose operations, respectively; $\left\|  \cdot  \right\|_2^2$ denotes Euclidean two norm of a vector; $diag\left(  \cdot  \right)$ represents a diagonal matrix; ${\bf{I}}_K$ is an $K\times K$ identity matrix.
%\textcolor[rgb]{0.00,0.00,1.00}{To enhance the readability of paper, the main acronyms are summarized in TABLE~\ref{Acronyms}.}

\section{Network Model}\label{Network Model}
\subsection{Network Descriptions}
\begin{figure}[t!]
    \begin{center}
        \includegraphics[width= 3.2in, height=2in]{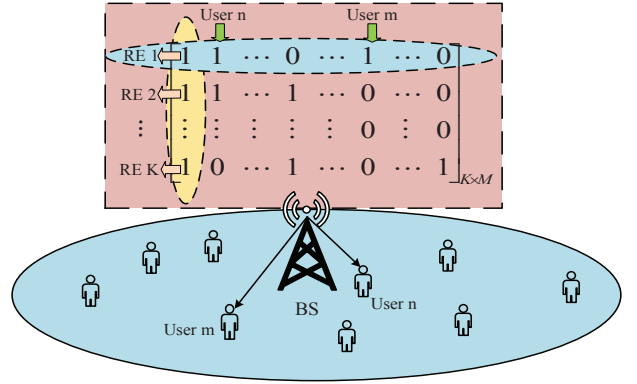}
         \caption{An illustration of the unified downlink NOMA transmission networks, where the spatial distributions of NOMA users follow homogeneous BPPs.}
        \label{PLS_NOMA_system_model}
    \end{center}
\end{figure}

As shown in Fig. \ref{PLS_NOMA_system_model}, we consider a unified downlink NOMA transmission scenario in a single cell\footnote{It is worth noting that estimating multi-cell scenarios in a unified NOMA framework can further enrich the contents of the paper considered \cite{Shin8010756}, which is set aside for our future work.}, where a base station (BS) transmits the information to $M$ randomly users. More precisely, the BS directly maps the data streams of multiple users into $K$ subcarriers or REs by utilizing one sparse spreading matrix ${{\bf{G}}_{K \times M}}$ (i.e, sparse matrix or codebook), in which there are a few number of non-zero entries within it and satisfies the relationship $M > K$.
To present straightforward results and analysis, we assume that the BS and NOMA users are equipped with a single antenna\footnote{Note that equipping multiple antennas on the nodes will further enhance the performance of CD/PD-NOMA networks, but this is beyond the scope of this treatise.}, respectively. Furthermore, assuming that the BS is located at the center of circular cluster denoted as $ {\cal D}$, with radius $R_{\cal D}$ and the spatial locations of $M$ users are modeled as homogeneous Binomial point processes (HBPPs) ${\Phi _l}$ \cite{Stochastic2012,Afshang7882710}.
To facilitate analysis, we assumed that $M$ users are divided into $M/2$ orthogonal pairs, in which distant user and nearby user can be distinguished based on their disparate channel conditions. Each pair of users is randomly selected to carry out NOMA protocol \cite{Ding2014performance}. A bounded path-loss model \cite{Stochastic2012} is employed to model the channel coefficients, which is capable of avoiding of singularity at small distances from the BS to users.
Meanwhile, these wireless links are disturbed by additive white Gaussian noise (AWGN) with mean power ${{N_0}}$.
Without loss of generality, the effective channel gains between the BS and users over subcarriers are sorted as $\left\| {{{\bf{h}}_M}} \right\|_2^2 >  \cdots  > \left\| {{{\bf{h}}_n}} \right\|_2^2 >  \cdots  > \left\| {{{\bf{h}}_m}} \right\|_2^2 >  \cdots  > \left\| {{{\bf{h}}_1}} \right\|_2^2$ \cite{David2003Order,Men7219393} with the assistance of order statistics.
In this treatise, we focus on the $m$-th user paired with the $n$-th user for NOMA transmission.
%\begin{table}[!t]
%\centering
%\caption{\textcolor[rgb]{0.00,0.00,1.00}{Table of Acronyms}}
%\tabcolsep5pt
%\renewcommand\arraystretch{1.1} % 调整表格高度
%\begin{tabular}{|c|c|}
%\hline
%\textbf{Terminologies}  &  \textbf{Acronyms} \\
%\hline
%Code-domain NOMA                    &  CD-NOMA  \\
%\hline
%Power-domain NOMA                    &  PD-NOMA  \\
%\hline
%Perfect successive interference cancellation    & pSIC \\
%\hline
%Imperfect successive interference cancellation    &  ipSIC   \\
%\hline
%Connection outage probability                     &  COP  \\
%\hline
%Existing outage probability formulation &   EXF    \\
%\hline
%Alternative outage probability Formulation  &      ALF            \\
%\hline
%\end{tabular}
%\label{Acronyms}
%\end{table}

\subsection{Signal Model}
Regarding the unified NOMA transmission in downlink single cell scenario, the BS transmits the superposed signals to multiple users, where the data stream of each user is spread over one column of sparse matrix. Hence the observation ${{\bf{y}}_\varphi } = {[{y_{\varphi 1}}{y_{\varphi 2}} \cdots {y_{\varphi K}}]^T}$ at the $\varphi $-th user over $K$ subcarriers  is given by
\begin{align}\label{The received signal expression at the n and m-th user}
{{\bf{y}}_\varphi }{\rm{ = }}diag\left( {{{\bf{h}}_\varphi }} \right) ( {{{\bf{g}}_n}\sqrt {{P_s}{a_n}} {x_n} + {{\bf{g}}_m}\sqrt {{P_s}{a_m}} {x_m}} ) + {{\bf{n}}_\varphi },
\end{align}
where $\varphi  \in \left( {n,m} \right)$. $x_{n}$ and $x_{m}$ are supposed to be normalized unity power signals for the $n$-th and $m$-th users, respectively, i.e, $\mathbb{E}\{x_{n}^2\}= \mathbb{E}\{x_{m}^2\}=1$.
Assuming the fixed power allocation coefficients satisfy the condition that ${a_m} > {a_n}$ with $a_m + a_n = 1$, which is for fairness considerations. Note that optimal power allocation coefficients \cite{Choi7383267,Zhu7982784} are capable of enhancing the performance of NOMA networks, but it is beyond the scope of this paper. $P_{s}$ denotes the normalized transmission power at the BS, i.e., $P_{s} = 1$.
The sparse indicator vector of the $\varphi $-th user is denoted by ${{\bf{g}}_\varphi } = {[{g_{\varphi 1}}{g_{\varphi 2}} \cdots {g_{\varphi K}}]^T}$, which is one column of ${{\bf{G}}_{K \times M}}$. More specifically, ${g_{\varphi k}}$ is the subcarrier index, where ${g_{\varphi k}}=1$ and ${g_{\varphi k}}=0$ indicate whether the signals are mapped into the corresponding RE or not.
Let ${{\bf{h}}_\varphi } = {[{{\tilde h}_{\varphi 1}}{{\tilde h}_{\varphi 2}} \cdots {{\tilde h}_{\varphi K}}]^T}$ denotes the channel vector between the BS and $\varphi $-th user occupying $K$ subcarriers with ${{\tilde h}_{\varphi k}} = \frac{{\sqrt \eta  {h_{\varphi k}}}}{{\sqrt {1 + {d^\alpha }} }}$, where ${h_{\varphi k}} \sim {\cal C}{\cal N}\left( {0,1} \right)$ is Rayleigh fading channel gain between the BS and $\varphi $-th user occupying the $k$-th subcarrier.
Additionally, $\eta$ and $\alpha $ are the frequency dependent factor and path loss exponent, respectively. $d$ denotes the distance from BS to $\varphi $-th user. ${{\bf{n}}_\varphi } \sim {\cal C}{\cal N}\left( {0,{N_0}{{\bf{I}}_{K}}} \right)$ denotes AWGN.
It is worth noting that based on the number of subcarriers, this unified framework can be reduced into CD-NOMA\footnote{It is worth pointing out that applying multi-dimensional constellations \cite{Wei1989}, channel coding (i.e., Low-Density Parity-Check (LDPC) codes or Turbo codes) and iterative decoding are capable of providing shaping gain and coding gain, which we may include in our future work.}
($K\ne 1$) and PD-NOMA ($K=1$), respectively. In particular, when $K$ is set to be one, the data streams of multiple users are mapped into one subcarrier, which can also be selected as a benchmark for CD-NOMA in the following. %\cite{Xiao2015,Bao2017}

To maximize the output SNRs and diversity orders, we employ the maximal ratio combiner (MRC) \cite{Tse2005fundamentals} at the $\varphi $-th user over $K$ subcarriers. Note that using MRC is not optimal but with low computational complexity. Let ${{\mathbf{u}}_\varphi } = \frac{{{{\left( {diag\left( {{{\mathbf{h}}_\varphi }} \right){{\mathbf{g}}_\varphi }} \right)}^*}}}{{\left\| {diag\left( {{{\mathbf{h}}_\varphi }}\right){{\mathbf{g}}_\varphi }} \right\|}}$, and then the received signal at the $\varphi $-th user can be written as
\begin{align}\label{The further received signal expression at the n and m-th user}
{{\tilde y}_\varphi }{\text{ = }}{{\mathbf{u}}_\varphi }diag\left( {{{\mathbf{h}}_\varphi }} \right) ( {{{\mathbf{g}}_n}\sqrt {{P_s}{a_n}} {x_n} + {{\mathbf{g}}_m}\sqrt {{P_s}{a_m}} {x_m}} ) + {{\mathbf{u}}_\varphi }{{\mathbf{n}}_\varphi }.
\end{align}
On the basis of aforementioned assumptions, the signal-plus-interference-to-noise ratio (SINR) at the $n$-th user to detect the $m$-th user's signal $x_{m}$ is given by % the $n$-th user first detect information of the $m$-th user. Therefore,
\begin{align}\label{The SINR expression at the n-th user to detect the m-th user}
{\gamma _{n \to m}} = \frac{{\rho \left\| {diag\left( {{{\bf{h}}_n}} \right){{\bf{g}}_m}} \right\|_2^2{a_m}}}{{\rho \left\| {diag\left( {{{\bf{h}}_n}} \right){{\bf{g}}_n}} \right\|_2^2{a_n} + 1}},
\end{align}
where $\rho  = \frac{{{P_s}}}{{{N_0}}}$ denotes the transmit SNR. For the sake of brevity, it is assumed that
${{{\bf{g}}_m}}$ and ${{{\bf{g}}_n}}$ have the same column weights for ${{\bf{G}}_{K \times M}}$. The optimization design of sparse matrix and spread sequence is capable of enhancing the performance of the unified NOMA framework, but this is beyond scope of this treatise.

By applying SIC technologies~\cite{Ding2017Mag}, the SINR of the $n$-th user, who needs to decode the information of itself is given by
\begin{align}\label{the SINR expression at n-th user to detect itself with SIC}
{\gamma _n} = \frac{{\rho \left\| {diag\left( {{{\bf{h}}_n}} \right){{\bf{g}}_n}} \right\|_2^2{a_n}}}{{\varpi \rho \left\| {{{\bf{h}}_I}} \right\|_2^2 + 1}},
\end{align}
where $\varpi {\rm{ = 0}}$ and $\varpi {\rm{ = 1}}$ denote the pSIC and ipSIC operations, respectively. Note that
${{\bf{h}}_I} = {[{h_{I1}}{h_{I2}} \cdots {h_{IK}}]^T}$ denotes the RI channel vector at $K$ subcarriers with $ {h_{Ik}} \sim {\cal C}{\cal N}\left( {0,{\Omega _I}} \right)$.

On the other hand, the $n$-th user is not always first detect the information of the $m$-th user and then decode its own signal. At this moment, the $n$-th user will decode the message of itself by directly treating the $m$-th user as interference without carrying out SIC operation. In this case, the corresponding SINR can be expressed as
\begin{align}\label{the SINR expression at n-th user to detect itself directly}
{\gamma _{n \to n}} = \frac{{\rho \left\| {diag\left( {{{\bf{h}}_n}} \right){{\bf{g}}_n}} \right\|_2^2 {a_n}}}{{\rho \left\| {diag\left( {{{\bf{h}}_n}} \right){{\bf{g}}_m}} \right\|_2^2 {a_m} + 1}}.
\end{align}

The SINR of a typical cell at the $m$-th NOMA user to decode the information of itself can be expressed as
\begin{align}\label{SINR m}
{\gamma _m} = \frac{{\rho \left\| {diag\left( {{{\bf{h}}_m}} \right){{\bf{g}}_m}} \right\|_2^2{a_m}}}{{\rho \left\| {diag\left( {{{\bf{h}}_m}} \right){{\bf{g}}_n}} \right\|_2^2{a_n} + 1}}.
\end{align}

\subsection{Channel statistical properties}
%这里给出公式（3）和公式（4）的表达式 and  公式（5）和公式（6）
In this subsection, different channel statistical properties are derived under the unified NOMA frameworks \cite{Yuanwei2017TWC}, which can be used for deriving the COP in the following sections.
\begin{lemma}\label{CD-NOMA:lemma:gamma channel_CDF for far user}
Assuming $M$ users randomly distributed within the circular cluster, the CDF ${F_{{\gamma _m}}}$ of the $m$-th user is given by
\begin{align}\label{CD-NOMA:the CDF of SINR expression for far user}
 &{F_{{\gamma _m}}}\left( x \right) \approx  {\phi _m}\sum\limits_{p = 0}^{M - m} {{
 M - m \choose
 p
 }\frac{{{{\left( { - 1} \right)}^p}}}{{m + p}}} \left[ {\sum\limits_{u = 1}^U {{b_u}} } \right. \nonumber \\
  &\times {\left. {\left( {1 - {e^{ - \frac{{x{c_u}}}{{\eta \rho \left( {{a_m} - x{a_n}} \right)}}}}\sum\limits_{i = 0}^{K - 1} {\frac{1}{{i!}}{{\left( {\frac{{x{c_u}}}{{\eta \rho \left( {{a_m} - x{a_n}} \right)}}} \right)}^i}} } \right)} \right]^{m + p}} ,
\end{align}
where ${a_m} > x{a_n}$, ${\phi _m} = \frac{{M!}}{{\left( {M - m} \right)!\left( {m - 1} \right)!}}$, ${
 M - m \choose
 p
 } = \frac{{\left( {M - m} \right)!}}{{p!\left( {M - m - p} \right)!}}$, ${b_u} = \frac{\pi }{{2U}}\sqrt {1 - \theta _u^2} \left( {{\theta _u} + 1} \right)$, ${c_u}{\rm{ = }}1 + {\left( {\frac{{{R_D}}}{2}\left( {{\theta _u} + 1} \right)} \right)^\alpha }$, ${\theta _u} = \cos \left( {\frac{{2u - 1}}{{2U}}\pi } \right)$ and $U$ is a parameter to ensure a complexity-accuracy tradeoff.
 \begin{proof}
See Appendix~A.
\end{proof}
\end{lemma}

\begin{lemma}\label{CD-NOMA:lemma:gamma channel_CDF for near user with SIC}
Assuming $M$ users randomly distributed within the circular cluster, the CDF $F_{{\gamma _n}}^{ipSIC}$ of the $n$-th user with ipSIC is given in \eqref{CD-NOMA:the CDF of SINR expression for near user with ipSIC} at the top of next page, where $\varpi  = 1$.
\begin{proof}
See Appendix~B.
\end{proof}
\begin{figure*}[!t]
\normalsize
\begin{align}\label{CD-NOMA:the CDF of SINR expression for near user with ipSIC}
 F_{{\gamma _n}}^{ipSIC} \approx \frac{{{\phi _n}}}{{\left( {K - 1} \right){\rm{!}}\Omega _I^K}}\sum\limits_{p = 0}^{M - n} {{
 M - n \choose
 p
 }\frac{{{{\left( { - 1} \right)}^p}}}{{n + p}}} \int_0^\infty  {{y^{K - 1}}{e^{ - \frac{y}{{{\Omega _I}}}}}} {\left[ {\sum\limits_{u = 1}^U {{b_u}} \left( {1 - {e^{ - \frac{{x{c_u}\left( {\varpi \rho y + 1} \right)}}{{\eta \rho {a_n}}}}}\sum\limits_{i = 0}^{K - 1} {\frac{1}{{i!}}{{\left( {\frac{{x{c_u}\left( {\varpi \rho y + 1} \right)}}{{\eta \rho {a_n}}}} \right)}^i}} } \right)} \right]^{n + p}}dy ,
\end{align}
\hrulefill \vspace*{0pt}
\end{figure*}
\end{lemma}
Substituting $\varpi  = 0$ into \eqref{CD-NOMA:the CDF of SINR expression for near user with ipSIC}, the CDF $F_{{\gamma _n}}^{pSIC}$ of the $n$-th user with pSIC is given by
\begin{align}\label{CD-NOMA:the CDF of SINR expression for near user with pSIC}
 F_{{\gamma _n}}^{pSIC}\approx & {\phi _n}\sum\limits_{p = 0}^{M - n} {{
 M - n \choose
 p
 }\frac{{{{\left( { - 1} \right)}^p}}}{{n + p}}} \left[ {\sum\limits_{u = 1}^U {{b_u}} } \right. \nonumber \\
  &\times {\left. {\left( {1 - {e^{ - \frac{{x{c_u}}}{{\eta \rho {a_n}}}}}\sum\limits_{i = 0}^{K - 1} {\frac{1}{{i!}}{{\left( {\frac{{x{c_u}}}{{\eta \rho {a_n}}}} \right)}^i}} } \right)} \right]^{n + p}}  .
\end{align}

\section{Performance evaluation}\label{Connection Outage}
Since the capacity of channel from the BS to the goal-directed user is less than the target transmission rate, the connection outage will occur \cite{Zhou5934342}. As a consequence, the goal-directed user is incapable of detecting the information accurately. In this section, the COP is selected as a metric to evaluate the performance of unified downlink NOMA networks. More specially, a pair of NOMA users (i.e., the $m$-th user and $n$-th user) for CD/PD-NOMA are characterized in terms of outage probabilities in the following.
\subsection{The COP of the $m$-th user}
The outage event of the $m$-th user in the typical cell is that the $m$-th user cannot detect its own information. Hence the COP of the $m$-th user for CD-NOMA can be expressed as
\begin{align}\label{CD-NOMA:the expression of COP for far user}
{P_{m,CD}} = {\rm{Pr}}\left( {{\gamma _m} < {\varepsilon _m}} \right),
\end{align}
where ${\varepsilon _m} = {2^{{R_m}}} - 1$ and $R_m$ is the target rate of the $m$-th user in the typical cell.

By applying \eqref{CD-NOMA:the CDF of SINR expression for far user}, the following theorem provides the COP of the $m$-th user.
\begin{theorem}\label{Theorem:CD-NOMA:the COP of far user}
The COP of the $m$-th user for CD-NOMA is given by
\begin{align}\label{CD-NOMA:the COP of far user}
{P_{m,CD}} \approx & {\phi _m}\sum\limits_{p = 0}^{M - m} {{
 M - m \choose
 p
 }\frac{{{{\left( { - 1} \right)}^p}}}{{m + p}}}  \nonumber \\
 & \times {\left[ {\sum\limits_{u = 1}^U {{b_u}} \left( {1 - {e^{ - \frac{{\tau {c_u}}}{\eta }}}\sum\limits_{i = 0}^{K - 1} {\frac{1}{{i!}}{{\left( {\frac{{\tau {c_u}}}{\eta }} \right)}^i}} } \right)} \right]^{m + p}}  ,
\end{align}
where $\tau  = \frac{{{\varepsilon _m}}}{{\rho \left( {{a_m} - {\varepsilon _m}{a_n}} \right)}}$ with ${a_m} > {\varepsilon _m}{a_n}$.
\end{theorem}

\begin{corollary}\label{Corollary:PD-NOMA:the COP of far user}
For the special case with $K=1$, the COP of the $m$-th user for PD-NOMA is given by
\begin{align}\label{PD-NOMA:the COP of far user}
{P_{m,PD}} \approx {\phi _m}\sum\limits_{p = 0}^{M - m} {{
 M - m \choose
 p
}\frac{{{{\left( { - 1} \right)}^p}}}{{m + p}}} {\left[ {\sum\limits_{u = 1}^U {{b_u}} \left( {1 - {e^{ - \frac{{{{{ \tau }}} {c_u}}}{\eta }}}} \right)} \right]^{m + p}}.
\end{align}
\end{corollary}
\subsection{The COP of the $n$-th user}
\subsubsection{Existing Outage Probability Formulation}
Considering a two-user case, the $m$-th user and $n$-th user are paired together to perform NOMA protocol. The outage for the $n$-th user can happen in the following two cases \cite{Ding2014performance,Liu7445146SWIPT}:
%1) the $n$-th user can not decode the message of the $m$-th user; 2) The $n$-th user can decode the message of the $m$-th user, then carries out SIC operations, but can not decode the information of itself.
\begin{itemize}
  \item The $n$-th user cannot decode the message of the $m$-th user.
  \item The $n$-th user can decode the message of the $m$-th user, then carries out SIC operations, but cannot decode the information of itself.
\end{itemize}
Based the aforementioned descriptions, the COP of the $n$-th user for existing formulation (EXF) can be expressed as
\begin{align}\label{CD-NOMA:the expression of COP for near user}
P_{n1,CD} =& \Pr \left\{ {{\gamma _{n \to m}} \le {\varepsilon _m}} \right\} \nonumber \\
&+ \Pr \left\{ {{\gamma _{n \to m}} > {\varepsilon _m},{\gamma _n} \le {\varepsilon _n}} \right\},
\end{align}
where ${\varepsilon _n} = {2^{{R_n}}} - 1$ with ${R_n}$ being the target rate at the $n$-th user to detect the $m$-th user.

The following theorem provides the COP of the $n$-th user with ipSIC for the downlink CD-NOMA networks.
\begin{theorem}\label{Theorem:CD-NOMA:the COP of near user for Case1 with ipSIC}
The COP of the $n$-th user with ipSIC for EXF in CD-NOMA networks is given by \eqref{CD-NOMA:the COP of near user for Case1 with ipSIC}, where $\beta {\text{ = }}\frac{{{\varepsilon _n}}}{{\rho {a_n}}}$, $\vartheta {\text{ = }}\frac{{\varpi {\varepsilon _n}}}{{{a_n}}}$ and $\varpi {\text{ = }}1$.
\begin{figure*}[!t]
\normalsize
\begin{align}\label{CD-NOMA:the COP of near user for Case1 with ipSIC}
P_{n1,CD}^{ipSIC} \approx \frac{{{\phi _n}}}{{\left( {K - 1} \right) {\rm{!}}\Omega _I^K}}\sum\limits_{p = 0}^{M - n} {{
   {M - n}  \choose
   p
} \frac{{{{\left( { - 1} \right)}^p}}}{{n + p}}} \int_0^\infty  {{y^{K - 1}}{e^{ - \frac{y}{{{\Omega _I}}}}}} {\left[ {\sum\limits_{u = 1}^U {{b_u}} \left( {1 - {e^{ - \frac{{{c_u}\left( {\vartheta y + \beta } \right)}}{\eta }}}\sum\limits_{i = 0}^{K - 1} {\frac{1}{{i!}}{{\left( {\frac{{\left( {\vartheta y + \beta } \right){c_u}}}{\eta }} \right)}^i}} } \right)} \right]^{n + p}}dy.
\end{align}
\hrulefill \vspace*{0pt}
\end{figure*}
\begin{proof}
See Appendix~C.
\end{proof}
\end{theorem}

Substituting $\varpi {\text{ = }}0$ into \eqref{CD-NOMA:the COP of near user for Case1 with ipSIC}, the COP of the $n$-th user with pSIC for EXF in CD-NOMA networks is given by
\begin{align}\label{CD-NOMA:the COP of near user for Case1 with pSIC}
P_{n1,CD}^{pSIC} \approx & {\phi _n}\sum\limits_{p = 0}^{M - n} {{
   {M - n}  \choose
   p
 }\frac{{{{\left( { - 1} \right)}^p}}}
{{n + p}}\left[ {\sum\limits_{u = 1}^U {{b_u}} } \right.}  \hfill  \nonumber \\
  &{\left. { \times \left( {1 - {e^{ - \frac{{\beta {c_u}}}
{\eta }}}\sum\limits_{i = 0}^{K - 1} {\frac{1}
{{i!}}{{\left( {\frac{{\beta {c_u}}}
{\eta }} \right)}^i}} } \right)} \right]^{n + p}} \hfill .
\end{align}

\begin{corollary}\label{Corollary:PD-NOMA:the COP of near user for case1}
For the special case with $K=1$, the COP of the $n$-th user with ipSIC for EXF in PD-NOMA networks is given by
\begin{align}\label{PD-NOMA :the COP of near user with ipSIC for Case1}
P_{n1,PD}^{ipSIC} \approx& \frac{{{\phi _n}}}{{{\Omega _I}}}\sum\limits_{p = 0}^{M - n} {{
   {M - n}  \choose
   p
}\frac{{{{\left( { - 1} \right)}^p}}}{{n + p}}} \nonumber  \\
&  \times \int_0^\infty  {{e^{ - \frac{y}{{{\Omega _I}}}}}} {\left[ {\sum\limits_{u = 1}^U {{b_u}} \left( {1 - {e^{ - \frac{{{c_u}\left( {\vartheta y + \beta } \right)}}{\eta }}}} \right)} \right]^{n + p}}dy  .
\end{align}
\end{corollary}

Substituting $\varpi {\text{ = }}0$ into \eqref{PD-NOMA :the COP of near user with ipSIC for Case1}, the COP of the $n$-th user with pSIC for EXF in PD-NOMA networks is given by
\begin{align}\label{PD-NOMA:the COP of near user with pSIC for Case1}
P_{n1,PD}^{pSIC} \approx & {\phi _n}\sum\limits_{p = 0}^{M - n} {{
   {M - n}  \choose
   p
 }\frac{{{{\left( { - 1} \right)}^p}}}
{{n + p}}\left[ {\sum\limits_{u = 1}^U {{b_u}} } \right.}  \hfill \nonumber \\
  & {\left. { \times \left( {1 - {e^{ - \frac{{ \beta {c_u}}}
{\eta }}}} \right)} \right]^{n + p}} \hfill .
\end{align}
\subsubsection{Alternative Outage Probability Formulation}
However, for the first case, when the decoding process for the $m$-th user at the $n$-th user fails, the outage event is not necessarily happened. Because the $n$-th user can still try to decode the message of itself by treating the $m$-th user's signal as interference without carrying out SIC operations. In other words, the previous outage formulation makes the decoding procedure of the $n$-th user highly depend the target rate of the $m$-th user, which ignores one possible case which can also support reliable transmission. As such, the alternative outage for the $n$-th user can happen in the following two cases:
\begin{itemize}
  \item The $n$-th user can not decode the message of the $m$-th user and the message of itself with treating the  $m$-th user's signal as interference.
  \item The $n$-th user can decode the message of the $m$-th user, but cannot detect the information of itself after carrying out SIC operations.
\end{itemize}
By the virtue of previous assumptions, the COP of the $n$-th user for alternative formulation (ALF) can be expressed as
\begin{align}\label{OP n}
P_{n2,CD} = &\Pr \left\{ {{\gamma _{n \to m}} \le {\varepsilon _m},{\gamma _{n \to n}} \le {\varepsilon _n}} \right\} \nonumber\\
&+ \Pr \left\{ {{\gamma _{n \to m}} > {\varepsilon _m},{\gamma _n} \le {\varepsilon _n}} \right\}.
\end{align}

The following theorem provides the COP of the $n$-th user with ipSIC for the downlink CD-NOMA networks.
\begin{theorem}\label{Theorem:CD-NOMA:the COP of near user with ipSIC for Case2}
The COP of the $n$-th user with ipSIC for ALF in CD-NOMA networks is given by  \eqref{CD-NOMA:the COP of near user with ipSIC for Case2}, where $\upsilon  = \frac{{{\varepsilon _n}}}
{{\rho \left( {{a_n} - {\varepsilon _n}{a_m}} \right)}}$ with ${a_n} > {\varepsilon _n}{a_m}$, $\zeta  = \min \left( {\tau ,\upsilon } \right)$.
\begin{figure*}[!t]
\normalsize
\begin{align}\label{CD-NOMA:the COP of near user with ipSIC for Case2}
  P_{n2,CD}^{ipSIC} \approx & {\phi _n}\sum\limits_{p = 0}^{M - n} {{
   {M - n}  \choose
   p
 }\frac{{{{\left( { - 1} \right)}^p}}}
{{n + p}}} {\left[ {\sum\limits_{u = 1}^U {{b_u}} \left( {1 - {e^{ - \frac{{\zeta {c_u}}}
{\eta }}}\sum\limits_{i = 0}^{K - 1} {\frac{1}
{{i!}}{{\left( {\frac{{\zeta {c_u}}}
{\eta }} \right)}^i}} } \right)} \right]^{n + p}} +
 \frac{{{\phi _n}}}{{\left( {K - 1} \right){\rm{!}}\Omega _I^K}} \nonumber \\
  & \times \sum\limits_{p = 0}^{M - n} {{
   {M - n}  \choose
   p
} \frac{{{{\left( { - 1} \right)}^p}}}{{n + p}}} \int_0^\infty  {{y^{K - 1}}{e^{ - \frac{y}{{{\Omega _I}}}}}} {\left[ {\sum\limits_{u = 1}^U {{b_u}} \left( {1 - {e^{ - \frac{{{c_u}\left( {\vartheta y + \beta } \right)}}{\eta }}}\sum\limits_{w = 0}^{K - 1} {\frac{1}{{w!}}{{\left( {\frac{{\left( {\vartheta y + \beta } \right){c_u}}}{\eta }} \right)}^w}} } \right)} \right]^{n + p}}dy \nonumber \\
&  - {\phi _n}\sum\limits_{p = 0}^{M - n} {{
   {M - n}  \choose
   p
}\frac{{{{\left( { - 1} \right)}^p}}}
{{n + p}}} {\left[ {\sum\limits_{u = 1}^U {{b_u}} \left( {1 - {e^{ - \frac{{\tau {c_u}}}
{\eta }}}\sum\limits_{i = 0}^{K - 1} {\frac{1}
{{i!}}{{\left( {\frac{{\tau {c_u}}}
{\eta }} \right)}^i}} } \right)} \right]^{n + p}} \hfill .
\end{align}
\hrulefill \vspace*{0pt}
\end{figure*}
\begin{proof}
See Appendix~D.
\end{proof}
\end{theorem}

Substituting $\varpi {\text{ = }}0$ into \eqref{CD-NOMA:the COP of near user with ipSIC for Case2}, the COP of the $n$-th user with pSIC for ALF in CD-NOMA networks is given by \eqref{CD-NOMA:the COP of near user with pSIC for Case2}.
\begin{figure*}[!t]
\normalsize
\begin{align}\label{CD-NOMA:the COP of near user with pSIC for Case2}
P_{n2,CD}^{pSIC} \approx &{\phi _n}\sum\limits_{p = 0}^{M - n} {{
   {M - n}  \choose
   p
 }\frac{{{{\left( { - 1} \right)}^p}}}
{{n + p}}} \left\{ {{{\left[ {\sum\limits_{u = 1}^U {{b_u}} \left( {1 - {e^{ - \frac{{\zeta {c_u}}}
{\eta }}}\sum\limits_{i = 0}^{K - 1} {\frac{1}
{{i!}}{{\left( {\frac{{\zeta {c_u}}}
{\eta }} \right)}^i}} } \right)} \right]}^{n + p}} + \left[ {\sum\limits_{u = 1}^U {{b_u}} } \right.} \right. \hfill\nonumber \\
 & \left. {{{\left. { \times \left( {1 - {e^{ - \frac{{\beta {c_u}}}
{\eta }}}\sum\limits_{i = 0}^{K - 1} {\frac{1}
{{i!}}{{\left( {\frac{{\beta {c_u}}}
{\eta }} \right)}^i}} } \right)} \right]}^{n + p}} - {{\left[ {\sum\limits_{u = 1}^U {{b_u}} \left( {1 - {e^{ - \frac{{\tau {c_u}}}
{\eta }}}\sum\limits_{i = 0}^{K - 1} {\frac{1}
{{i!}}{{\left( {\frac{{\tau {c_u}}}
{\eta }} \right)}^i}} } \right)} \right]}^{n + p}}} \right\} \hfill .
\end{align}
\hrulefill \vspace*{0pt}
\end{figure*}

\begin{corollary}\label{Corollary:PD-NOMA:the COP of near user for case2}
For the special case with $K=1$, the COP of the $n$-th user with ipSIC for ALF in PD-NOMA networks is given by \eqref{PD-NOMA :the COP of near user with ipSIC for Case2}.
\begin{figure*}[!t]
\normalsize
\begin{align}\label{PD-NOMA :the COP of near user with ipSIC for Case2}
 P_{n2,PD}^{ipSIC} \approx &{\phi _n}\sum\limits_{p = 0}^{M - n} {{
   {M - n}  \choose
   p
}  \frac{{{{\left( { - 1} \right)}^p}}}{{n + p}}} \left\{ {{{\left[ {\sum\limits_{u = 1}^U {{b_u}\left( {1 - {e^{ - \frac{{\varsigma {c_u}}}{\eta }}}} \right)} } \right]}^{n + p}} - {{\left[ {\sum\limits_{u = 1}^U {{b_u}\left( {1 - {e^{ - \frac{{\tau {c_u}}}{\eta }}}} \right)} } \right]}^{n + p}}} \right\} \nonumber \\
 \begin{array}{*{20}{c}}
   {} & {} & {}  \\
\end{array} & + \frac{{{\phi _n}}}{{{\Omega _I}}}\sum\limits_{p = 0}^{M - n} {
   {M - n}  \choose
   p
} \frac{{{{\left( { - 1} \right)}^p}}}{{n + p}}\int_0^\infty  {{e^{ - \frac{y}{{{\Omega _I}}}}}} {\left[ {\sum\limits_{u = 1}^U {{b_u}\left( {1 - {e^{ - \frac{{{c_u}\left( {\vartheta y + \beta } \right)}}{\eta }}}} \right)} } \right]^{n + p}}dy
\end{align}
\hrulefill \vspace*{0pt}
\end{figure*}
\end{corollary}

Substituting $\varpi {\text{ = }}0$ into \eqref{PD-NOMA :the COP of near user with ipSIC for Case2}, the COP of the $n$-th user with pSIC for ALF in PD-NOMA networks is given by
\begin{align}\label{PD-NOMA:the COP of near user with pSIC for Case2}
&P_{n2,PD}^{pSIC} \approx \sum\limits_{p = 0}^{M - n} {{
   {M - n}  \choose
   p
 }} \frac{{{{\left( { - 1} \right)}^p}{\phi _n}}}
{{n + p}}{\left\{ {\left[ {\sum\limits_{u = 1}^U {{b_u}} \left( {1 - {e^{ - \frac{{ \zeta {c_u}}}
{\eta }}}} \right)} \right]} \right.^{n + p}} \hfill \nonumber \\
 & \left. { + {{\left[ {\sum\limits_{u = 1}^U {{b_u}} \left( {1 - {e^{ - \frac{{ \beta {c_u}}}
{\eta }}}} \right)} \right]}^{n + p}} - {{\left[ {\sum\limits_{u = 1}^U {{b_u}} \left( {1 - {e^{ - \frac{{ \tau {c_u}}}
{\eta }}}} \right)} \right]}^{n + p}}} \right\} \hfill .
\end{align}

\begin{proposition}\label{proposition:The COP of the selected user pair}
The COP of the selected user pairing with ipSIC/pSIC for CD/PD-NOMA are given by
\begin{align}\label{the COP of the selected user pair for CD-NOMA}
P_{nm,CD}^\psi  = 1 - \left( {1 - {P_{m,CD}}} \right)\left( {1 - P_{{\tilde n},CD}^\psi } \right),
\end{align}
and
\begin{align}\label{the asymptotic OP of x1 with ipSIC}
P_{nm,PD}^\psi  = 1 - \left( {1 - {P_{m,PD}}} \right)\left( {1 - P_{{\tilde n},PD}^\psi } \right),
\end{align}
respectively, where $\psi  \in \left( {ip{\rm{SIC}},pSIC} \right)$ and ${\tilde n}\in \left( {n1,n2} \right)$. ${P_{m,CD}}$ and ${P_{m,PD}}$ are given by \eqref{CD-NOMA:the COP of far user} and \eqref{PD-NOMA:the COP of far user}, respectively. $P_{{n1},CD}^{ipSIC}$, $P_{{n1},PD}^{ipSIC}$, $P_{{n1},CD}^{pSIC}$ and $P_{{n1},PD}^{pSIC}$ are given by \eqref{CD-NOMA:the COP of near user for Case1 with ipSIC}, \eqref{CD-NOMA:the COP of near user for Case1 with pSIC}, \eqref{PD-NOMA :the COP of near user with ipSIC for Case1} and  \eqref{PD-NOMA:the COP of near user with pSIC for Case1}, respectively. $P_{{n2},CD}^{ipSIC}$, $P_{{n2},PD}^{ipSIC}$, $P_{{n2},CD}^{pSIC}$ and $P_{{n2},PD}^{pSIC}$ are given by \eqref{CD-NOMA:the COP of near user with ipSIC for Case2}, \eqref{CD-NOMA:the COP of near user with pSIC for Case2}, \eqref{PD-NOMA :the COP of near user with ipSIC for Case2} and \eqref{PD-NOMA:the COP of near user with pSIC for Case2}, respectively.
\end{proposition}

\subsection{Diversity Order Analysis}
To gain more deep insights, diversity order is usually selected to be a matric to evaluate the system performance, which highlights the slope of the curves for outage probabilities varying with the SNRs. Hence the definition of diversity order is given by
\begin{align}\label{The diversity order of COP}
d =  - \mathop {\lim }\limits_{\rho  \to \infty } \frac{{\log \left( {P^\infty }\left( \rho  \right) \right)}}{{\log \rho }},
\end{align}
where ${P^\infty }\left( \rho  \right)$ denotes the asymptotic COP.

\begin{corollary}\label{Corollary:the expression of asymptotic COP for the m-th user}
Based on analytical result in \eqref{CD-NOMA:the COP of far user}, the asymptotic COP of the $m$-th user at high SNR for CD-NOMA is given by
\begin{align}\label{The expression of asymptotic COP for the m-th user with CD-NOMA}
P_{m,CD}^\infty  \approx \frac{{M!}}
{{\left( {M - m} \right)!m!}}{\left[ {\sum\limits_{u = 1}^U  \frac{{{b_u}}}
{{K!}}{{\left( {\frac{{\tau {c_u}}}
{\eta }} \right)}^K}} \right]^m}  \propto  \frac{1}{{{\rho ^{mK}}}},
\end{align}
where $ \propto $ represents ``be proportional to".
\begin{proof}
To facilitate the calculation, define the summation term in \eqref{CD-NOMA:the COP of far user}, i.e., ${\Theta _1}{\text{ = }}1 - {e^{ - \frac{{\tau {c_u}}}{\eta }}}\underbrace {\sum\limits_{i = 0}^{K - 1} {\frac{1}
{{i!}}{{\left( {\frac{{\tau {c_u}}}{\eta }} \right)}^i}} }_{{\Theta _2}}$. Applying power series expansion, the summation term ${\Theta _2}$ can be rewritten as ${\Theta _2} = {e^{\frac{{\tau {c_u}}} {\eta }}} - \sum\limits_{i = K}^\infty  {\frac{1}
{{i!}}{{\left( {\frac{{\tau {c_u}}}{\eta }} \right)}^i}} $.  Substituting ${\Theta _2}$ into ${\Theta _1}$, when $x  \to 0 $, ${\Theta _1}$ with the approximation of ${e^{ - x}} \approx 1 - x$ is formulated as ${\Theta _1} \approx \frac{1}{{K!}}{\left( {\frac{{\tau {c_u}}}{\eta }} \right)^K}$. As a further development, substituting ${\Theta _1}$ into \eqref{CD-NOMA:the COP of far user} and taking the first term $(p=0)$ \cite{Men7454773}, we obtain \eqref{The expression of asymptotic COP for the m-th user with CD-NOMA}. Obviously, $P_{m,CD}^\infty $ is  a function of $\rho$, which is proportional to $\frac{1}{{{\rho ^{mK}}}}$. The proof is completed.
\end{proof}
\end{corollary}

For the special case with $K=1$, the asymptotic COP of the $m$-th user at high SNR for PD-NOMA is given by
\begin{align}\label{The expression of asymptotic COP for the m-th user with PD-NOMA}
P_{m,PD}^\infty  \approx \frac{{M!}}
{{\left( {M - m} \right)!m!}}{\left[ {\sum\limits_{u = 1}^U {{b_u}} \left( {\frac{{\tau {c_u}}}
{\eta }} \right)} \right]^m} \propto  \frac{1}{{{\rho ^{m}}}}.
\end{align}
\begin{remark}\label{remark1: the m-th user}
Upon substituting \eqref{The expression of asymptotic COP for the m-th user with CD-NOMA} and \eqref{The expression of asymptotic COP for the m-th user with PD-NOMA} into \eqref{The diversity order of COP}, the diversity orders of the $m$-th user for CD-NOMA and PD-NOMA are $mK$ and $m$, respectively.% CD-NOMA 与 subcarriers 有关
\end{remark}

\begin{corollary}\label{Corollary:the expression of asymptotic COP for the n-th user CD-NOMA case1}
Based on analytical result in \eqref{CD-NOMA:the COP of near user for Case1 with ipSIC}, when $\rho  \to \infty$, the asymptotic COP of the $n$-th user with ipSIC for EXF in CD-NOMA networks is given by
\begin{align}\label{The asymptotic expression of the n-th user with ipSIC for CD-NOMA case1}
P_{n1,CD}^{ipSIC,\infty } \approx \frac{{{\phi _n}}}{{\left( {K - 1} \right){\rm{!}} \Omega _I^K}}\sum\limits_{p = 0}^{M - n} {{
   {M - n}  \choose
   p
}\frac{{{{\left( { - 1} \right)}^p}}}{{n + p}}} \int_0^\infty  {{y^{K - 1}}}  \nonumber \\
 \times {e^{ - \frac{y}{{{\Omega _I}}}}}{\left[ {\sum\limits_{u = 1}^U {{b_u}} \left( {1 - {e^{ - \frac{{y\vartheta {c_u}}}{\eta }}}\sum\limits_{i = 0}^{K - 1} {\frac{1}{{i!}}{{\left( {\frac{{y\vartheta {c_u}}}{\eta }} \right)}^i}} } \right)} \right]^{n + p}}dy.
\end{align}
\end{corollary}

Substituting $\varpi  = 0$ into \eqref{The asymptotic expression of the n-th user with ipSIC for CD-NOMA case1}, the asymptotic COP of the $n$-th user with pSIC at high SNR for EXF  in CD-NOMA networks is given by
\begin{align}\label{The asymptotic expression of the n-th user with pSIC for CD-NOMA case1}
P_{n1,CD}^{pSIC,\infty } \approx \frac{{M!}}
{{\left( {M - n} \right)!n!}}{\left[ {\sum\limits_{u = 1}^U {\frac{{{b_u}}}
{{K!}}} {{\left( {\frac{{\beta {c_u}}}
{\eta }} \right)}^K}} \right]^n} \propto  \frac{1}{{{\rho ^{nK}}}}.
\end{align}

\begin{remark}\label{remark2: the n-th user for CD-NOMA in case1}
Upon substituting \eqref{The asymptotic expression of the n-th user with ipSIC for CD-NOMA case1} and \eqref{The asymptotic expression of the n-th user with pSIC for CD-NOMA case1} into \eqref{The diversity order of COP}, the diversity orders of the $n$-th user with ipSIC/pSIC for EXF in CD-NOMA networks are zero and $nK$, respectively.
\end{remark}

\begin{corollary}\label{Corollary:the expression of asymptotic COP for the n-th user PD-NOMA case1}
For the special case with $K=1$ in \eqref{The asymptotic expression of the n-th user with ipSIC for CD-NOMA case1}, the asymptotic COP of the $n$-th user with ipSIC at high SNR for EXF in PD-NOMA networks is given by
\begin{align}\label{The asymptotic expression of the n-th user with ipSIC for PD-NOMA case1}
 P_{n1,PD}^{ipSIC,\infty } \approx &\frac{{{\phi _n}}}{{{\Omega _I}}}\sum\limits_{p = 0}^{M - n} {{
   {M - n}  \choose
   p
}\frac{{{{\left( { - 1} \right)}^p}}}{{n + p}}}  \nonumber \\
  &\times \int_0^\infty  {{e^{ - \frac{y}{{{\Omega _I}}}}}} {\left[ {\sum\limits_{u = 1}^U {{b_u}} \left( {1 - {e^{ - \frac{{y\vartheta {c_u}}}{\eta }}}} \right)} \right]^{n + p}}dy .
\end{align}
\end{corollary}
Substituting $\varpi  = 0$ into \eqref{The asymptotic expression of the n-th user with ipSIC for PD-NOMA case1}, the asymptotic COP of the $n$-th user at high SNR with pSIC in PD-NOMA networks for EXF is given
\begin{align}\label{The asymptotic expression of the n-th user with pSIC for PD-NOMA case1}
P_{n1,PD}^{pSIC,\infty } \approx \frac{{M!}}
{{\left( {M - n} \right)!n!}}{\left[ {\sum\limits_{u = 0}^U {{b_u}} \left( {\frac{{\tau {c_u}}}
{\eta }} \right)} \right]^n} \propto  \frac{1}{{{\rho ^{n}}}}.
\end{align}
\begin{remark}\label{remark3: the n-th user for PD-NOMA in case1}
Upon substituting \eqref{The asymptotic expression of the n-th user with ipSIC for PD-NOMA case1} and \eqref{The asymptotic expression of the n-th user with pSIC for PD-NOMA case1} into \eqref{The diversity order of COP}, the diversity orders of the $n$-th user with ipSIC/pSIC for EXF in PD-NOMA networks are zero and $n$, respectively.
\end{remark}

\begin{corollary}\label{Corollary7:the expression of asymptotic COP for the n-th user CD-NOMA case2}
The asymptotic COP of the $n$-th user with ipSIC at high SNR for ALF in CD-NOMA networks is given by \eqref{The asymptotic expression of the n-th user with ipSIC for CD-NOMA case2} at the top of next page.
\begin{figure*}[!t]
\normalsize
\begin{align}\label{The asymptotic expression of the n-th user with ipSIC for CD-NOMA case2}
P_{n2,CD}^{ipSIC,\infty } \approx & \frac{{M!}}{{\left( {M - n} \right)!n!}}{\left[ {\sum\limits_{u = 1}^U {\frac{{{b_u}}}{{K!}}} {{\left( {\frac{{\zeta {c_u}}}{\eta }} \right)}^K}} \right]^n} - \frac{{M!}}{{\left( {M - n} \right)!n!}}{\left[ {\sum\limits_{u = 1}^U {\frac{{{b_u}}}{{K!}}} {{\left( {\frac{{\tau {c_u}}}{\eta }} \right)}^K}} \right]^n}{\rm{ + }}\frac{{{\phi _n}}}{{\left( {K - 1} \right){\rm{!}} \Omega _I^K}} \nonumber \\
 & \times \sum\limits_{p = 0}^{M - n} {{
   {M - n}  \choose
   p
}\frac{{{{\left( { - 1} \right)}^p}}}{{n + p}}} \int_0^\infty  {{y^{K - 1}}{e^{ - \frac{y}{{{\Omega _I}}}}}} {\left[ {\sum\limits_{u = 1}^U {{b_u}} \left( {1 - {e^{ - \frac{{y\vartheta {c_u}}}{\eta }}}\sum\limits_{w = 0}^{K - 1} {\frac{1}{{w!}}{{\left( {\frac{{y\vartheta {c_u}}}{\eta }} \right)}^w}} } \right)} \right]^{n + p}}dy.
\end{align}
\hrulefill \vspace*{0pt}
\end{figure*}
\end{corollary}

Similar to the solving process of \eqref{The asymptotic expression of the n-th user with pSIC for CD-NOMA case1}, substituting $\varpi  = 0$ into \eqref{The asymptotic expression of the n-th user with ipSIC for CD-NOMA case2}, the asymptotic COP of the $n$-th user at high SNR with pSIC in CD-NOMA networks for ALF is given by
\begin{align}\label{The asymptotic expression of the n-th user with pSIC for CD-NOMA case2}
  &P_{n2,CD}^{pSIC,\infty }  \approx  \frac{{M!}}
{{\left( {M - n} \right)!n!}}\left\{ {{{\left[ {\sum\limits_{u = 1}^U {\frac{{{b_u}}}
{{K!}}{{\left( {\frac{{\zeta {c_u}}}
{\eta }} \right)}^K}} } \right]}^n}} \right. \hfill  \nonumber \\
  & \left. { + {{\left[ {\sum\limits_{u = 1}^U {\frac{{{b_u}}}
{{K!}}{{\left( {\frac{{\beta {c_u}}}
{\eta }} \right)}^K}} } \right]}^n} - {{\left[ {\sum\limits_{u = 1}^U {\frac{{{b_u}}}
{{K!}}{{\left( {\frac{{\tau {c_u}}}
{\eta }} \right)}^K}} } \right]}^n}} \right\} \hfill \propto  \frac{1}{{{\rho ^{nK}}}}.
\end{align}
\begin{remark}\label{remark4: the n-th user for CD-NOMA in case2}
Upon substituting \eqref{The asymptotic expression of the n-th user with ipSIC for CD-NOMA case2} and \eqref{The asymptotic expression of the n-th user with pSIC for CD-NOMA case2} into \eqref{The diversity order of COP}, the diversity orders of the $n$-th user with ipSIC/pSIC for ALF in CD-NOMA networks are zero and $nK$, respectively.
\end{remark}

\begin{corollary}\label{Corollary8:the expression of asymptotic COP for the n-th user PD-NOMA case2}
For the special case with $K=1$, the asymptotic COP of the $n$-th user at high SNR  for ALF in PD-NOMA networks is given by \eqref{The asymptotic expression of the n-th user with ipSIC for PD-NOMA case2}  at the top of next page.
\begin{figure*}[!t]
\normalsize
\begin{align}\label{The asymptotic expression of the n-th user with ipSIC for PD-NOMA case2}
P_{n2,PD}^{ipSIC,\infty } \approx &\frac{{M!}}{{\left( {M - n} \right)!n!}}{\left[ {\sum\limits_{u = 1}^U {{b_u}\left( {\frac{{\zeta {c_u}}}{\eta }} \right)} } \right]^n} - \frac{{M!}}{{\left( {M - n} \right)!n!}}{\left[ {\sum\limits_{u = 1}^U {{b_u}\left( {\frac{{\tau {c_u}}}{\eta }} \right)} } \right]^n} \nonumber \\
  & + \frac{{{\phi _n}}}{{{\Omega _I}}}\sum\limits_{p = 0}^{M - n} {{
   {M - n}  \choose
   p
}\frac{{{{\left( { - 1} \right)}^p}}}{{n + p}}} \int_0^\infty  {{e^{ - \frac{y}{{{\Omega _I}}}}}} {\left[ {\sum\limits_{u = 1}^U {{b_u}} \left( {1 - {e^{ - \frac{{y\vartheta {c_u}}}{\eta }}}} \right)} \right]^{n + p}}dy .
\end{align}
\hrulefill \vspace*{0pt}
\end{figure*}
\end{corollary}
Substituting $\varpi  = 0$ into \eqref{The asymptotic expression of the n-th user with ipSIC for PD-NOMA case2}, the asymptotic COP of the $n$-th user at high SNR with pSIC for ALF in PD-NOMA networks is given
\begin{align}\label{The asymptotic expression of the n-th user with pSIC for PD-NOMA case2}
&P_{n2,PD}^{pSIC,\infty }\approx   \frac{{M!}}
{{\left( {M - n} \right)!n!}}\left\{ {{{\left[ {\sum\limits_{u = 1}^U {{b_u}} \left( {\frac{{\xi {c_u}}}
{\eta }} \right)} \right]}^n}} \right. \hfill \nonumber \\
 & \left. { + {{\left[ {\sum\limits_{u = 1}^U {{b_u}} \left( {\frac{{\beta {c_u}}}
{\eta }} \right)} \right]}^n} - {{\left[ {\sum\limits_{u = 1}^U {{b_u}} \left( {\frac{{\tau {c_u}}}
{\eta }} \right)} \right]}^n}} \right\} \hfill \propto  \frac{1}{{{\rho ^{n}}}}.
\end{align}

As can be seen from \eqref{The asymptotic expression of the n-th user with ipSIC for PD-NOMA case2}, the third term of $P_{n2,PD}^{ipSIC,\infty }$ is a constant and leads directly to the fact that $P_{n2,PD}^{ipSIC,\infty }$ is not proportional to $ \frac{1}{{{\rho }}}$.
However, in \eqref{The asymptotic expression of the n-th user with pSIC for PD-NOMA case2}, $P_{n2,PD}^{pSIC,\infty }$ is proportional to $\frac{1}{{{\rho ^{n}}}}$. Hence the observation can be obtained in the following.
\begin{remark}\label{remark5: the n-th user for PD-NOMA in case2}
Upon substituting \eqref{The asymptotic expression of the n-th user with ipSIC for PD-NOMA case2} and \eqref{The asymptotic expression of the n-th user with pSIC for PD-NOMA case2} into \eqref{The diversity order of COP}, the diversity orders of the $n$-th user with ipSIC/pSIC for ALF in PD-NOMA networks are zero and $n$, respectively.
\end{remark}

From the above remarks, we can observe that CD-NOMA with pSIC is capable of providing a higher diversity order than PD-NOMA. Hence we can adjust the size of subcarriers $K$ to meet different application requirements. Additionally, we find that due to the impact of RI, CD/PD-NOMA with ipSIC obtain zero diversity order. The design of an efficient SIC is important for NOMA networks.

\begin{remark}\label{Remark_Add}
Under the condition of fixed SNR, the outage probabilities are determined by the SIC used for interference cancellation as well as the number of subcarriers and the size of target rate.
\end{remark}

\begin{proposition}\label{proposition:The asymptotic COP of the selected user pair}
The asymptotic COP of the selected user pairing with ipSIC/pSIC for CD/PD-NOMA at high SNR are given by
\begin{align}\label{the asymptotic COP of the selected user pair for CD-NOMA}
P_{nm,CD}^{\psi ,\infty } = P_{m,CD}^\infty  + P_{\tilde n,CD}^{\psi ,\infty } - P_{m,CD}^\infty P_{\tilde n,CD}^{\psi ,\infty },
\end{align}
and
\begin{align}\label{the asymptotic asymptotic OP of x1 with ipSIC}
P_{nm,PD}^{\psi ,\infty } = P_{m,PD}^\infty  + P_{\tilde n,PD}^{\psi ,\infty } - P_{m,PD}^\infty P_{\tilde n,PD}^{\psi ,\infty },
\end{align}
respectively, where $P_{m,CD}^\infty $ and $P_{m,PD}^\infty $ are given by \eqref{The expression of asymptotic COP for the m-th user with CD-NOMA} and \eqref{The expression of asymptotic COP for the m-th user with PD-NOMA}, respectively. $P_{n1,CD}^{ipSIC ,\infty }$, $P_{n1,CD}^{pSIC ,\infty }$, $P_{n1,PD}^{ipSIC ,\infty }$ and $P_{n1,PD}^{pSIC ,\infty }$ are given by \eqref{The asymptotic expression of the n-th user with ipSIC for CD-NOMA case1}, \eqref{The asymptotic expression of the n-th user with pSIC for CD-NOMA case1}, \eqref{The asymptotic expression of the n-th user with ipSIC for PD-NOMA case1}  and \eqref{The asymptotic expression of the n-th user with pSIC for PD-NOMA case1}, respectively. $P_{n2,CD}^{ipSIC ,\infty }$, $P_{n2,CD}^{pSIC ,\infty }$, $P_{n2,PD}^{ipSIC ,\infty }$ and $P_{n2,PD}^{pSIC ,\infty }$ are given by \eqref{The asymptotic expression of the n-th user with ipSIC for CD-NOMA case2}, \eqref{The asymptotic expression of the n-th user with pSIC for CD-NOMA case2} and \eqref{The asymptotic expression of the n-th user with ipSIC for PD-NOMA case2} and \eqref{The asymptotic expression of the n-th user with pSIC for PD-NOMA case2}, respectively.
\end{proposition}
\begin{remark}\label{remark6: user pairing for CD/PD-NOMA with ipSIC/pSIC}
On the basis of conclusions of above corollaries, the diversity orders of the selected user pairing with ipSIC/pSIC for CD-NOMA and PD-NOMA are zero/$mK$ and zero/$m$, respectively. As can be observed that due to the impact of RI for imperfect cancellation process,
the selected user pairing with ipSIC for CD/PD-NOMA obtain zero diversity order. Additionally, it is shown that the diversity orders of the selected user pairing are determined by the $m$-th user.
\end{remark}

As shown in TABLE~\ref{Summarize}, the relationship between different factors for CD/PD-NOMA, such as, outage probability formulations, SIC schemes and diversity order, are summarized to illustrate the comparison between them.
In TABLE~\ref{Summarize}, we use ``F'' , ``S'' and ``D'' to represent outage probability formulation, SIC scheme and diversity order, respectively.
\begin{table}[!h]
\begin{center}
\caption{The relationship between different factors for CD/PD NOMA networks.}
{\tabcolsep9.5pt %设置表格左右宽度
\renewcommand\arraystretch{1} %设置表格上下高度
\begin{tabular}{|c|c|c|c|c|}\hline   %{p{3cm}p{3cmp}p{3cm}p{3cm}}%
 %% \diagbox[width=3em,trim=l]{}   &\diagbox[width=3em,trim=l]{}  &\diagbox[width=3em,trim=l]{} & D & S \\
  \textbf{Pairing users} & \textbf{NOMA}  &\textbf{F} & \textbf{S} & \textbf{D} \\
     \hline
\multirow{4}{*}{The $m$-th user}    & \multirow{2}{*}{CD-NOMA}   & ------ & ------ & \multirow{2}{*}{$mK$}   \\
\cline{3-4}
                     &  &  ------  &  ------ &   \\
\cline{2-5}
                     & \multirow{2}{*}{PD-NOMA} & ------   & ------    &  \multirow{2}{*}{$m$} \\
\cline{3-4}
                &        &  ------  & ------   & \\
\hline
\multirow{8}{*}{The $n$-th user}  & \multirow{4}{*}{CD-NOMA}  &  \multirow{2}{*}{EXF}  &  ipSIC    & $ 0 $ \\
\cline{4-5}
                                  &                           &                       &     pSIC   & $ nK   $ \\
\cline{3-5}
               &\multirow{8}{*}{PD-NOMA}  &   \multirow{2}{*}{ALF}   &  ipSIC       & $0$  \\
\cline{4-5}
                                  &                           &                       &     pSIC   & $ nK   $ \\
\cline{2-5}
               &     &\multirow{2}{*}{EXF}     &  ipSIC     & $  0$ \\
\cline{4-5}
               &     &     &   pSIC    & $n$ \\
\cline{3-5}
               &     &\multirow{2}{*}{ALF}     & ipSIC     & $  0$ \\
\cline{4-5}
               &     &     &   pSIC    & $n$ \\
\hline
\end{tabular}}{}
\label{Summarize}
\end{center}
\end{table}

\subsection{Throughput Analysis}\label{Throughput Analysis}
In this subsection, the system throughput of the unified NOMA framework is characterized in delay-limited transmission mode.
In this mode, the BS transmits information at a constant rate $R$, which is subject to the effect of outage probability due to wireless fading channels.
\paragraph{CD-NOMA case}
According to the analytical results derived in the above section, using \eqref{CD-NOMA:the COP of far user}, \eqref{CD-NOMA:the COP of near user for Case1 with ipSIC} and \eqref{CD-NOMA:the COP of near user with ipSIC for Case2}, the system throughput of CD-NOMA with ipSIC/pSIC is given by
\begin{align}\label{Delay-limited Transmission for CD-NOMA}
R_{CD} ^{\psi} = \left( {1 - {P_{m,CD}}} \right){R_n} + \left( {1 - P_{\tilde n,CD}^\psi } \right){R_m}.
\end{align}
\paragraph{PD-NOMA case}
Similar to \eqref{Delay-limited Transmission for CD-NOMA}, using \eqref{PD-NOMA:the COP of far user}, \eqref{CD-NOMA:the COP of near user for Case1 with pSIC} and \eqref{PD-NOMA:the COP of near user with pSIC for Case2}, the system throughput of PD-NOMA with ipSIC/pSIC is given by
\begin{align}\label{Delay-limited Transmission direct for PD-NOMA}
R_{PD} ^{\psi} = \left( {1 - {P_{m,PD}}} \right){R_n} + \left( {1 - P_{\tilde n,PD}^\psi} \right){R_m}.
\end{align}

%\textcolor[rgb]{0.00,0.00,1.00}{
%\begin{remark}\label{Remark_Add_ADD}
%Under the condition of fixed SNR,
%\end{remark}
%}

\section{Numerical Results}\label{Numerical Results}
In this section, we focus on investigating a typical pair of users with random pairing.
Monte Carlo simulation parameters used in this section are summarized in TABLE~\ref{parameter} \cite{Liu7445146SWIPT,Ding6623110}, where BPCU is short for bit per channel use and the pass loss exponent $\alpha =2 $ aims to simplify simulation analysis.
The complexity-vs-accuracy tradeoff parameter is set to be $U=15$ and simulation results are denoted by $ \bullet $. Additionally, the conventional OMA is selected to be a benchmark for comparison purposes. The target rate ${R_o}$ for OMA satisfies the relationship with ${R_o} = {R_n} + {R_m}$.
Note that the setting of smaller target data rate for NOMA users can be applied into the Internet of Things (IoT) scenarios, which require low energy consumption, small packet service and so on.
\begin{table}[!t]
\centering
\caption{Table of Parameters for Numerical Results}
\tabcolsep5pt
\renewcommand\arraystretch{1.1} % 调整表格高度
\begin{tabular}{|l|l|}
\hline
Monte Carlo simulations repeated  &  ${10^5}$ iterations \\
\hline
Carrier frequency  &  $1 $ GHz  \\
\hline
\multirow{1}{*}{Power allocation coefficients of NOMA} &  \multirow{1}{*}{ $a_m=0.8$, $a_n=0.2$}   \\
\hline
\multirow{1}{*}{Target data rates}  & \multirow{1}{*}{$R_{{n}}=R_{{m}}=0.01 $ BPCU}  \\
\cline{1-2}
Pass loss exponent  & $\alpha=2$  \\
\hline
The radius of the user zone  &  $R_D=2$ m \\
\hline
\end{tabular}
\label{parameter}
\end{table}

\begin{figure}[t!]
    \begin{center}
        \includegraphics[width=3.48in,  height=2.8in]{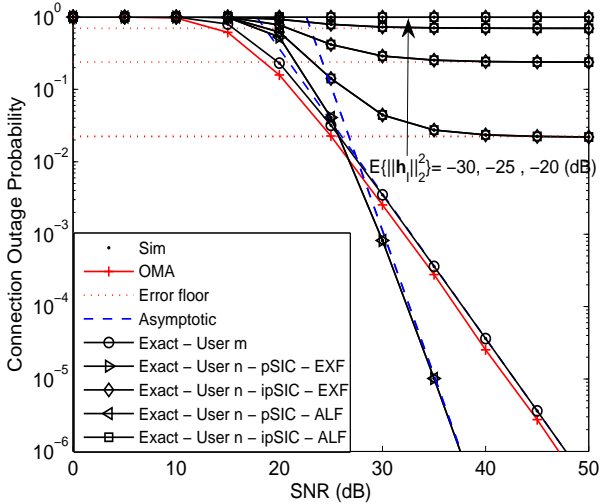}
        \caption{The COP versus the transmit SNR, with $K=2$, $M=3$, $n=2$, $m=1$, $R_D$ = 2 m, $R_n$ = $R_m$ = 0.01 BPCU.}
        \label{The_COP_pSIC_ipSIC_n_m_diff_LI}
    \end{center}
\end{figure}

Fig. \ref{The_COP_pSIC_ipSIC_n_m_diff_LI} plots the COP of a pair of NOMA users (the $m$-th and $n$-th user) versus the transmit SNR with ipSIC/pSIC, where $K = 2$. In particular, the different values of RI are set to be from $-30$ dB to $-20$ dB.
The exact analytical curve for the outage probability of the $m$-th user is plotted according to \eqref{CD-NOMA:the COP of far user}. Furthermore, the exact analytical curves for the outage probability of the $n$-th user with both ipSIC and pSIC for EXF and ALF are plotted based on \eqref{CD-NOMA:the COP of near user for Case1 with ipSIC}, \eqref{CD-NOMA:the COP of near user with ipSIC for Case2} and \eqref{CD-NOMA:the COP of near user for Case1 with pSIC}, \eqref{CD-NOMA:the COP of near user with pSIC for Case2}, respectively. Obviously, the exact outage probability curves match perfectly with Monte Carlo simulations results. It is observed that
the outage performance of OMA is inferior to the $n$-th user with pSIC and superior to the $m$-th user. This is due to the fact that NOMA is also capable of providing better fairness since multiple users are served simultaneously, which is the same as the conclusions in \cite{Ding2014performance,Yuexinwei7812773}.
Additionally, as can be observed from figure, the dashed curves represent the asymptotic COP of the $m$-th user and $n$-th user with pSIC for EXF and ALF, which can be obtained by numerically evaluating \eqref{The expression of asymptotic COP for the m-th user with CD-NOMA}, \eqref{The asymptotic expression of the n-th user with pSIC for CD-NOMA case1} and \eqref{The asymptotic expression of the n-th user with pSIC for CD-NOMA case2}. One can observe that the asymptotic outage probabilities are approximated to the analytical results in the high SNR regime. The dotted curves represent the asymptotic outage probabilities of the $n$-th user with ipSIC for EXF and ALF, which are calculated from \eqref{The asymptotic expression of the n-th user with ipSIC for CD-NOMA case1} and \eqref{The asymptotic expression of the n-th user with ipSIC for CD-NOMA case2}, respectively. It is shown that the outage performance of the $n$-th user with ipSIC converges to an error floor and obtain zero diversity order, which verifies the insights in \textbf{Remark \ref{remark2: the n-th user for CD-NOMA in case1}} and \textbf{Remark \ref{remark3: the n-th user for PD-NOMA in case1}}. Due to the influence of RI, the outage behavior of the $n$-th user with ipSIC is inferior to OMA. The reason is that the RI signal from imperfect cancellation operation is the dominant impact factor.
With the value of RI increasing from $-30$ dB to $-20$ dB, the outage behavior of the $n$-th user is becoming more worse and deteriorating. More specifically, when $\mathbb{E} \{ {\left\| {{{\bf{h}}_I}} \right\|_2^2} \}= -10$ dB, the outage probability of the $n$-th user will be always one. Hence the design of effective multiuser receiver algorithm is significant to improve the performance of NOMA networks in practical scenario.

\begin{figure}[t!]
    \begin{center}
        \includegraphics[width=3.48in,  height=2.8in]{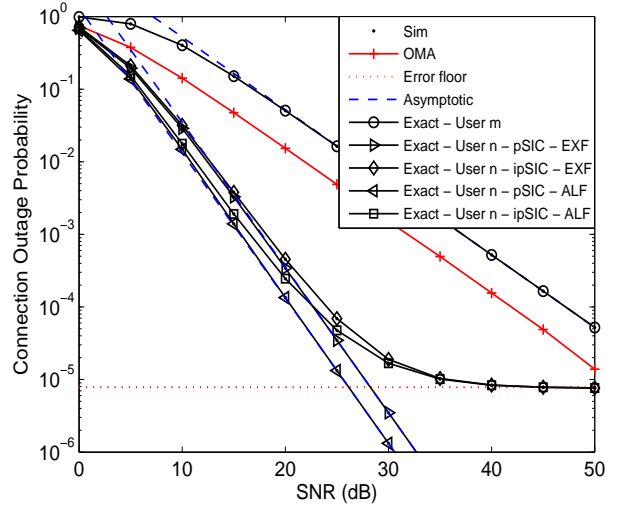}
        \caption{The COP versus the transmit SNR, with $M=3$, $n=2$, $m=1$, $K = 1$, $R_n=0.1$, $R_m=0.5$ BPCU, $\mathbb{E} \{ {\left\| {{{\bf{h}}_I}} \right\|_2^2} \}= -30$ dB.}
        \label{The_OP_M3_n2_m1_K1_RD2_unequal_TR}
    \end{center}
\end{figure}

Another important observation can be seen from Fig. \ref{The_COP_pSIC_ipSIC_n_m_diff_LI}, when the target rates of the $n$-th user for EXF and ALF are set to be equal, there are the identical outage probabilities. This is because when the target rate of $n$-th user is greater than or equal to that of the $m$-th user, the $n$-th user with ALF forcibly decode the message of itself without carrying out SIC operations, which will be seriously constrained by interference from the $m$-th user.
Another scenario is that when the target rate of the $n$-th user is less than that of the $m$-th user, which will be discussed in the following.

\begin{figure}[t!]
    \begin{center}
        \includegraphics[width=3.48in,  height=2.8in]{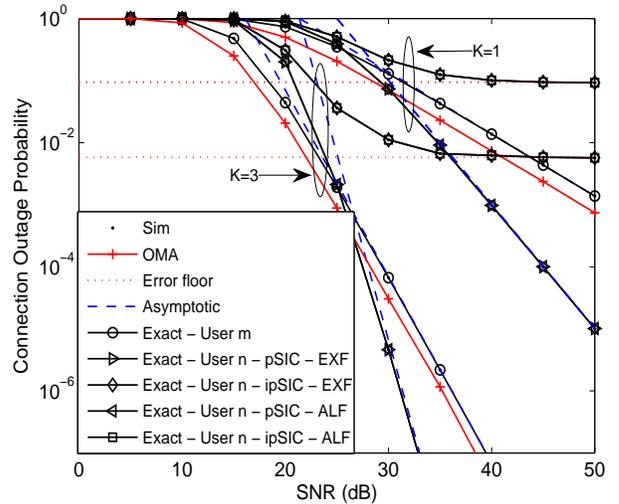}
        \caption{The COP versus the transmit SNR, with $M=3$, $n=2$, $m=1$, $R_n$ = $R_m$ = 0.01 BPCU, $\mathbb{E} \{ {\left\| {{{\bf{h}}_I}} \right\|_2^2} \}= -30$ dB.}
        \label{The_OP_m_n_pSIC_ipSIC_n_m_diff_K_add_OMA}
    \end{center}
\end{figure}

As a further advance, Fig. \ref{The_OP_M3_n2_m1_K1_RD2_unequal_TR} plots the COP versus SNR with unequal target rates for user pairing.
More particularly, the target rates of the $n$-th user and $m$-th user satisfy the relationship $R_n < R_m$. It is observed that the outage behavior of the $n$-th user with ALP has an advantage over that of the $n$-th user with EXF. Under this assumption, the $n$-th user tries to detect its own information without carrying out the SIC operation and it will suffer from less interference.
Another observation is that the outage behavior of the $n$-th user with ipSIC also precede OMA and converge to the same error floor in the high SNR region. This behavior is caused by the fact that the RI signal is not the dominant factor.

Fig. \ref{The_OP_m_n_pSIC_ipSIC_n_m_diff_K_add_OMA} plots the COP versus SNR with the different number of subcarriers (i.e., $K=3$ and $K=1$). For the special case with $K=1$, the unified NOMA framework is reduced into PD-NOMA.
The exact outage probability curve of the $m$-th user for PD-NOMA is plotted according to \eqref{PD-NOMA:the COP of far user}. The exact outage probability curves of the $n$-th user with ipSIC and pSIC for both EXF and ALF
are given by Monte Carlo simulations and precisely match with the analytical expressions which have been derived in \eqref{PD-NOMA :the COP of near user with ipSIC for Case1}, \eqref{PD-NOMA:the COP of near user with pSIC for Case1}, \eqref{PD-NOMA :the COP of near user with ipSIC for Case2} and \eqref{PD-NOMA:the COP of near user with pSIC for Case2}, respectively.
As can be observed from figure, the asymptotic outage probabilities of this pair of users for PD-NOMA are also approximated with the analytical results in the high SNR regime. As can be observed that with the subcarriers $K$ increasing, CD-NOMA has a more steep slope and provide better outage performance relative to PD-NOMA. This is due to the fact that CD-NOMA is capable of achieving the higher diversity orders, which verify the conclusion in \textbf{Remark \ref{remark4: the n-th user for CD-NOMA in case2}} and \textbf{Remark \ref{Remark_Add}}. Hence we can confirm that the diversity gains of CD-NOMA are directly combined with the number of subcarriers.

\begin{figure}[t!]
    \begin{center}
        \includegraphics[width=3.48in,  height=2.8in]{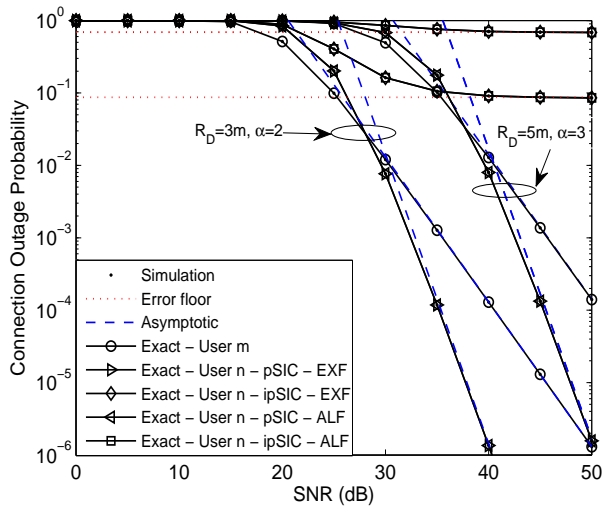}
        \caption{The COP versus the transmit SNR, with $K=2$, $M=3$, $n=2$, $m=1$, $R_n$ = $R_m$ = 0.01 BPCU, $\mathbb{E} \{ {\left\| {{{\bf{h}}_I}} \right\|_2^2} \}= -30$ dB.}
        \label{The_OP_pSIC_ipSIC_n_m_diff_RD_new}
    \end{center}
\end{figure}
Fig. \ref{The_OP_pSIC_ipSIC_n_m_diff_RD_new} plots the COP versus SNR for different network radius and pass-loss factors. As can be observed that with the decreased network radius, the better outage behaviors of the selected user pairing can be obtained. This is due to the fact that a smaller network radius results in a lower path-loss. Similarly, if the pass-loss factor is adjusted from $\alpha=3$ to $\alpha=2$, the better outage performance can also be achieved. As a consequence, from a practical perspective, the design of NOMA systems should be in conjunction with cell radius and pass-loss exponent.
Additionally, the setting of target rates for the users is critical in NOMA networks.

\begin{figure}[t!]
    \begin{center}
        \includegraphics[width=3.48in,  height=2.8in]{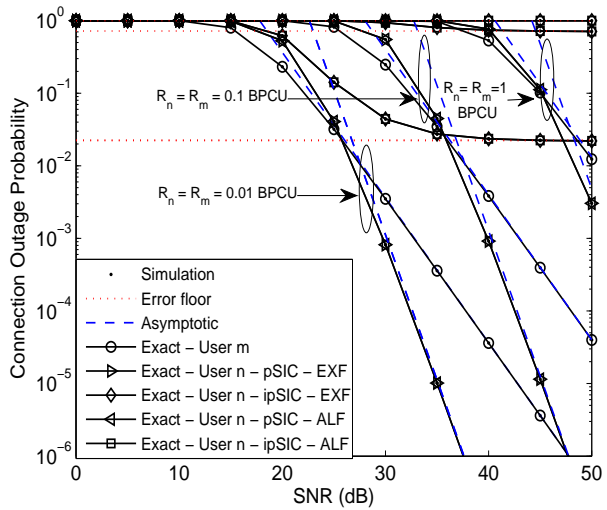}
        \caption{The COP versus the transmit SNR, with $M=3$, $n=2$, $m=1$, $K = 2$, $\mathbb{E} \{ {\left\| {{{\bf{h}}_I}} \right\|_2^2} \}= -30$ dB.}
        \label{The_OP_pSIC_ipSIC_n_m_diff_TR_new}
    \end{center}
\end{figure}

Fig. \ref{The_OP_pSIC_ipSIC_n_m_diff_TR_new} plots the outage probabilities versus SNR for different user target rates. We observe that as the target rate decreases, the lower outage probabilities can be achieved. This is due to the fact that the achievable rates are directly related to the target SNRs. It is beneficial to detect the superposed signals for the selected user pairing with smaller target SNRs. It is worth pointing out that the impact of practical scenario parameter frequency dependent factor $\eta$ has been taken into account in the unified NOMA framework. Furthermore, the incorrect choice of $R_n$ and $R_m$ will lead to the improper outage behavior for the unified framework, which versifies the conclusion in \textbf{Remark \ref{Remark_Add}}.

\begin{figure}[t!]
    \begin{center}
        \includegraphics[width=3.48in,  height=2.8in]{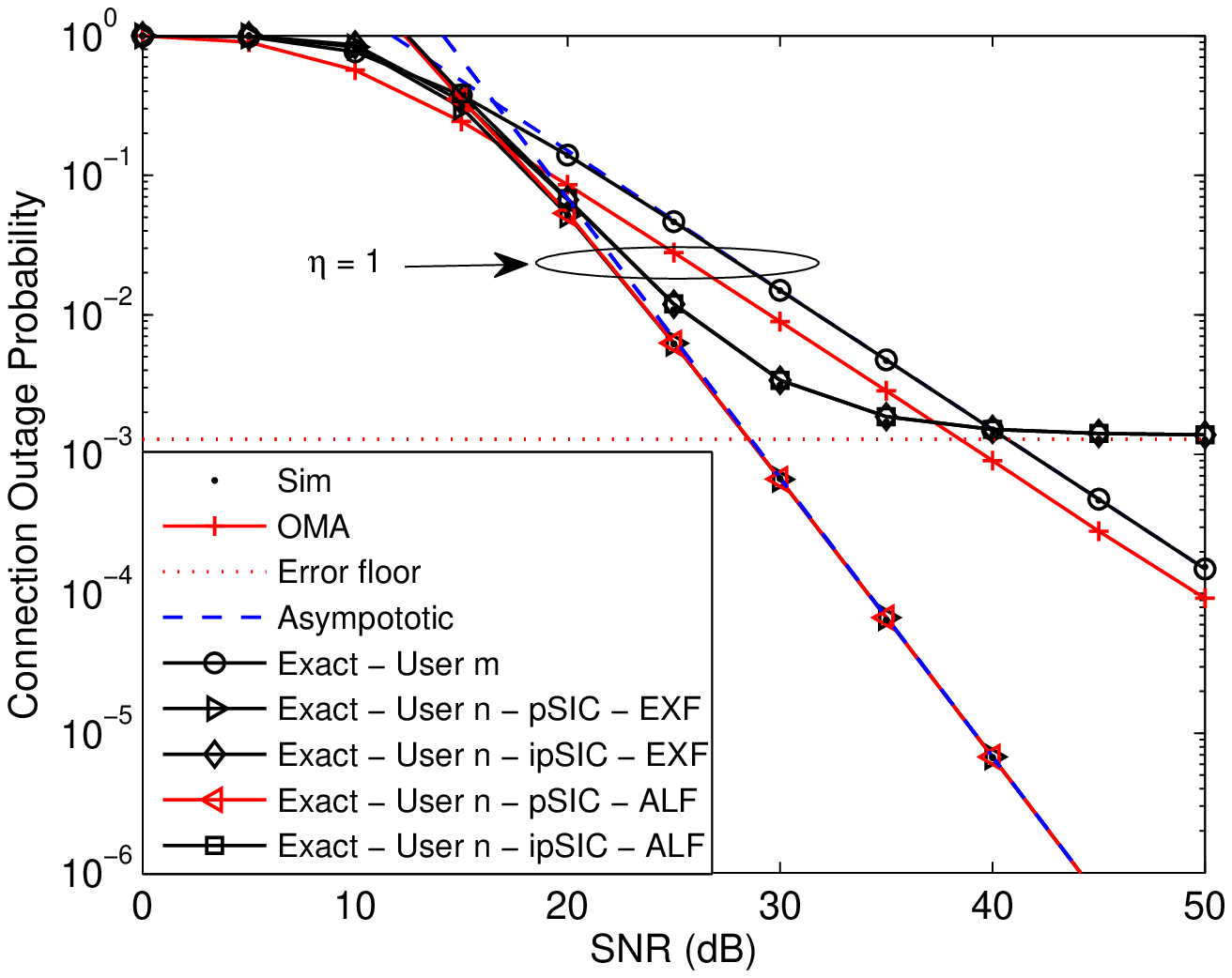}
        \caption{The COP versus the transmit SNR, with $M=3$, $n=2$, $m=1$, $K = 1$, $R_n=R_m=1$ BPCU, $\mathbb{E} \{ {\left\| {{{\bf{h}}_I}} \right\|_2^2} \}= -30$ dB.}
        \label{The_COP_pSIC_ipSIC_diff_dependant_fre}
    \end{center}
\end{figure}
\begin{figure}[t!]
    \begin{center}
        \includegraphics[width=3.48in,  height=2.8in]{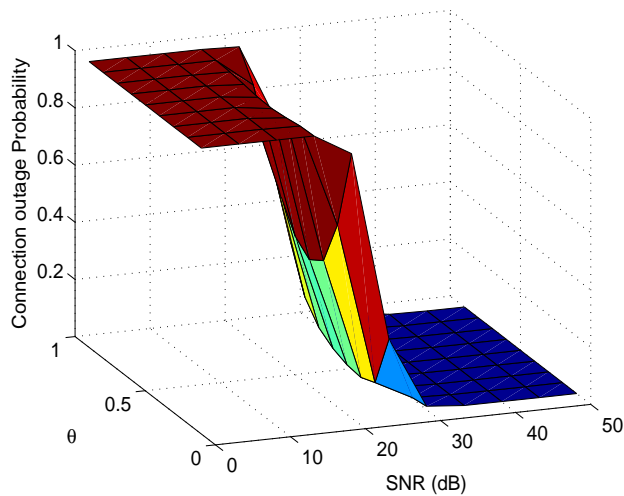}
        \caption{COP of the user pairing versus the transmit SNR and $\theta$, with $M=3$, $n=2$, $m=1$, $R_n$ = $R_m$ = 0.01 BPCU.}
        \label{The_OP_m_n_pSIC_3D}
    \end{center}
\end{figure}
In order to obtain tractable analytical results, Fig. \ref{The_COP_pSIC_ipSIC_diff_dependant_fre} plots the COP versus SNR with frequency dependent factor $\eta = 1$, where the target rates $R_n$ and $R_m$ for the user pair are set to be $R_n=R_m=1$ BPCU. As can be observed that the outage behavior of the $n$-th user with pSIC is also superior to that of the $m$-th user. This is due to the fact that we have $n>m$ and the $n$-th user achieves a higher diversity order. Another observation is that the outage behavior of the $n$-th user with ipSIC exceeds OMA on the condition of low SNR region. This is because RI is not the main influence factor in the low SNR region. This phenomenon indicates that it is significant to select favorable network parameter. Apparently, optimizing $\eta$ can enhance the network outage performance.

To illustrate the impact of dynamic power factor on NOMA performance, Fig. 8 plots the COP of the selected user pairing versus SNR and $\theta$, where $\theta  \in \left[ {0,1} \right]$ is the dynamic power allocation factor. Especially, when $a_n$ is set to be $a_n = \theta$, $a_m = 1 - \theta$.  The exact analytical results with pSIC are calculated from (23). One can observed that the COP decreases as SNR increases, which is the same as the traditional trend, where the COP always decreases as the transmit SNR increases. The reason is that the COP of the selected user pairing is determined by the $m$-th user. Another observation is that the dynamic power allocation factor affect the optimal COP with different values of SNR. This phenomenon indicates that it is significant to select beneficial system parameters. Furthermore, optimizing the power allocation factor is capable of further enhancing the COP.

\begin{figure}[t!]
    \begin{center}
        \includegraphics[width=3.48in,  height=2.8in]{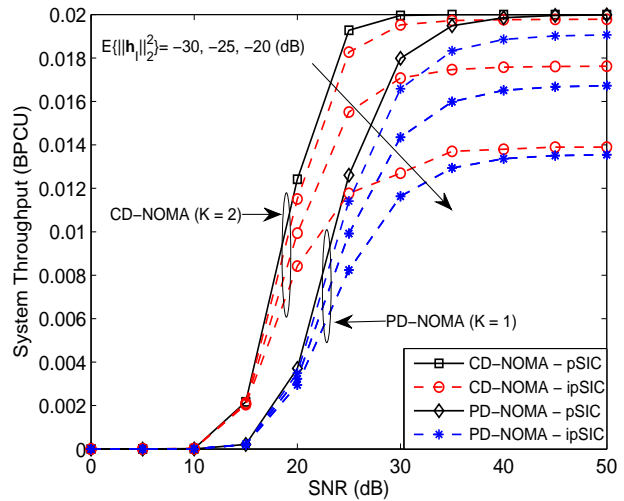}
        \caption{System throughput versus the transmit SNR, with $M=3$, $n=2$, $m=1$.}
        \label{The_delay_limited_throughput_diff_LI}
    \end{center}
\end{figure}
Fig. \ref{The_delay_limited_throughput_diff_LI} plots the system throughput versus SNR in the delay-limited transmission mode.
The solid black curves represent throughput of CD-NOMA and PD-NOMA with pSIC, which can be obtained from \eqref{Delay-limited Transmission for CD-NOMA} and \eqref{Delay-limited Transmission direct for PD-NOMA}, respectively.
The red and blue dash curves represent throughput of CD-NOMA and PD-NOMA with ipSIC for the different values of RI, respectively.
We observe that CD-NOMA attains a higher throughput compared to PD-NOMA, since CD-NOMA has the smaller outage probabilities. This is due to that CD-NOMA is capable of attaining the larger diversity order than that of PD-NOMA.
Another observation is that increasing the values of RI from $-30$ dB to $-20$ dB will reduce the system throughput in high SNR region. This is because that CD/PD-NOMA converge to the error floors in the high SNR region. In addition, it is worth noting that adjusting the size of target data rate (i.e., $R_{m}$ and $R_{n}$) will affect the system throughput in delay-limited transmission mode.

\section{Conclusions}\label{Conclusion}
In this paper, a unified downlink NOMA transmission scenario has been investigated insightfully. Stochastic geometry has been employed for modeling the locations of NOMA users in the networks. More specifically, the performance of the unified NOMA framework has been characterized in terms of COP. The exact expressions of outage probability for a pair of users with ipSIC/pSIC have been derived.
It was demonstrated the diversity orders of the $m$-th user for CD/PD-NOMA are $mK$ and $m$, respectively. However, due to the impact of RI, the diversity orders achieved by the $n$-th user with ipSIC are zeros for CD/PD-NOMA. On the basis of analytical results, we observed that the outage behaviors of CD-NOMA is superior to that of PD-NOMA. Additionally, it was shown that CD/PD-NOMA can provide better fairness than OMA when multiple users are served simultaneously. The system throughput of CD/PD-NOMA with ipSIC/pSIC for the unified framework has been discussed in delay-limited transmission mode. Numerical results were presented to verify the analysis. Applying the unified NOMA framework considered into other scenarios, i.e., cooperative communications, physical layer security and so on, are capable of further providing additional design insights, which is one promising future research direction.

\appendices
\section*{Appendix~A: Proof of Lemma \ref{CD-NOMA:lemma:gamma channel_CDF for far user}} \label{Appendix:A}
\renewcommand{\theequation}{A.\arabic{equation}}
\setcounter{equation}{0}

The proof starts by assuming ${{{\bf{g}}_m}}$ and ${{{\bf{g}}_n}}$ have the same column weights for ${{\bf{G}}_{K \times M}}$.
That is to say that ${\left\| {diag\left( {{{\bf{h}}_m}} \right){{\bf{g}}_m}} \right\|_2^2}$ and ${\left\| {diag\left( {{{\bf{h}}_m}} \right){{\bf{g}}_n}} \right\|_2^2}$ follow the same distribution. Based on \eqref{SINR m}, the expression of CDF ${F_{{\gamma _m}}}$ for the $m$-th user with ${a_m} > x{a_n}$ is given by
\begin{align}\label{The CDF of the m-th user}
{F_{{\gamma _m}}}\left( x \right){\rm{ = {\text{Pr}}}}\left( {Z_m < \frac{{x}}{{\rho \left( {{a_m} - x{a_n}} \right)}}}  \buildrel \Delta \over = z \right),
\end{align}
where $Z_m = \left\| {diag\left( {{{\bf{h}}_m}} \right){{\bf{g}}_m}} \right\|_2^2 = \frac{\eta }{{1 + {d^\alpha }}}\sum\limits_{k = 1}^K {{{\left| {{g_{mk}}{{\tilde h}_{mk}}} \right|}^2}}$. It is observed that $Y = \sum\limits_{k = 1}^K {{{\left| {{g_{mk}}{{\tilde h}_{mk}}} \right|}^2}} $ is subject to a Gamma distribution with the parameters of $\left( {K,1} \right)$. The corresponding CDF and PDF of ${Y}$ are given by
\begin{align}\label{the CDF for Gamma distrbution}
{F_Y}\left( y \right) = 1 - {e^{ - y}}\sum\limits_{i = 0}^{K - 1} {\frac{{{y^i}}}{{i!}}} ,
\end{align}
and
\begin{align}\label{the PDF for Gamma distrbution}
{f_Y}\left( y \right) = \frac{{{y^{K - 1}}{e^{ - y}}}}{{\left( {K - 1} \right){\rm{!}}}},
\end{align}
respectively.

On the basis of order statistics \cite{David2003Order}, the CDF ${F_{Z_m}}$ of the sorted channel gains between the BS and the $m$-th user over $K$ subcarriers has a specific relationship with the unsorted channel gain, which can be expressed as
\begin{align}\label{the CDF for Gamma distrbution}
{F_{Z_m}}\left( z \right) = {\phi _m}\sum\limits_{p = 0}^{M - m} {{
 M - m \choose
 p
 }\frac{{{{\left( { - 1} \right)}^p}}}{{m + p}}} {\left[ {F_{{{\tilde Z}_m}}}(z) \right]^{m + p}},
\end{align}
where ${F_{{{\tilde Z}_m}}}$ denotes the CDF of unsorted channels for the $m$-th user.
The next focus is to calculate the CDF ${F_{{{\tilde Z}_m}}}$. Recalling the hypothetical conditions, the density function of the distance can be found by using the fact that the locations of the users are uniformly distributed in ${\cal D}$ \cite{Stochastic2012,Ding6623110}. For any area $A$ with a size of $\Lambda$ and $A \in {\cal D}$, the distribution of each point $W$ is written as ${\mathop{\rm P}\nolimits} \left( {W \in A} \right) = \frac{\Lambda }{{\pi R_{\cal D}^2}}$ and the corresponding PDF is written as ${p_W}\left( w \right) = \frac{1}{{\pi R_D^2}}$. As a further development, the CDF ${F_{{{\tilde Z}_m}}}$ is given by
\begin{align}
 & {F_{{{{{\tilde Z}_m}}}}}\left( z \right) \nonumber \\
  & = \int\limits_D {\left[ {1 - {e^{ - \frac{{\left( {1 + {d^\alpha }} \right)z}}{\eta }}}\sum\limits_{i = 0}^{K - 1} {\frac{1}{{i!}}{{\left( {\frac{{\left( {1 + {d^\alpha }} \right)z}}{\eta }} \right)}^i}} } \right]} {p_W}\left( w \right)dw ,
\end{align}
where $d$ is determined by the distance between the point $W$ and BS.
By applying polar coordinate conversion, ${F_{{{\tilde Z}_m}}}(z)$ can be further given by
\begin{align}\label{the CDF of Z for unsorted channels with stocastic}
&{F_{{{\tilde Z}_m}}}(z)  \nonumber \\
&= \frac{2}{{R_D^2}}\int_0^{{R_D}} {\left[ {1 - {e^{ - \frac{{z\left( {1 + {r^\alpha }} \right)}}{\eta }}}\sum\limits_{i = 0}^{K - 1} {\frac{1}{{i!}}{{\left( {\frac{{z\left( {1 + {r^\alpha }} \right)}}{\eta }} \right)}^i}} } \right]} rdr.
\end{align}

For an arbitrary choice of $\alpha $, it is difficult to obtain effective insights from the integral in (A.6). Hence we use the Gaussian-Chebyshev quadrature \cite{Hildebrand1987introduction} to provide an approximation of \eqref{the CDF of Z for unsorted channels with stocastic} and rewrite it as
\begin{align}\label{the simplified CDF of Z for unsorted channels with stocastic}
{F_{{{\tilde Z}_m}}}(z) \approx \sum\limits_{u = 1}^U {{b_u}} \left( {1 - {e^{ - \frac{{z {c_u}}}{\eta }}}\sum\limits_{i = 0}^{K - 1} {\frac{1}{{i!}}{{\left( {\frac{{z {c_u}}}{\eta }} \right)}^i}} } \right).
\end{align}

Substituting \eqref{the simplified CDF of Z for unsorted channels with stocastic} into \eqref{the CDF for Gamma distrbution}, we can obtain \eqref{CD-NOMA:the CDF of SINR expression for far user}. The proof is completed.

%the CDF ${F_{Z_m}}$ of sorted channel gains is given by
%\begin{align}\label{the simplified CDF of Z for sorted channels with stocastic}
% {F_{Z_m}}\left( z \right) \approx& {\phi _m}\sum\limits_{p = 0}^{M - m} {{
% M - m \choose
% p
% }\frac{{{{\left( { - 1} \right)}^p}}}{{m + p}}}  \nonumber \\
%  &\times {\left[ {\sum\limits_{u = 1}^U {{b_u}} \left( {1 - {e^{ - \frac{{z{c_u}}}{\eta }}}\sum\limits_{i = 0}^{K - 1} {\frac{1}{{i!}}{{\left( {\frac{{z{c_u}}}{\eta }} \right)}^i}} } \right)} \right]^{m + p}}.
%\end{align}
%Finally, substituting $z = \frac{{x}}{{\rho \left( {{a_m} - x{a_n}} \right)}}$ into \eqref{the simplified CDF of Z for sorted channels with stocastic}, we can obtain \eqref{CD-NOMA:the CDF of SINR expression for far user}. The proof is completed.

\appendices
\section*{Appendix~B: Proof of Lemma \ref{CD-NOMA:lemma:gamma channel_CDF for near user with SIC}}\label{Appendix:B}
\renewcommand{\theequation}{B.\arabic{equation}}
\setcounter{equation}{0}

Based on \eqref{the SINR expression at n-th user to detect itself with SIC}, the CDF of ${F_{{\gamma _n}}}$ is formulated as
\begin{align}\label{The CDF of the n-th user}
 {F_{{\gamma _n}}}\left( x \right) &= {{\text{Pr}}}\left( {{\gamma _n} < x} \right) \nonumber \\
   &  = {{\text{Pr}}}\left( {\frac{{\rho {a_n}\left\| {diag\left( {{{\bf{h}}_n}} \right){{\bf{g}}_n}} \right\|_2^2}}{{\varpi \rho \left\| {{{\bf{h}}_I}} \right\|_2^2 + 1}} < x} \right),
\end{align}
where $\varpi =1$. Denoting ${Z_n} = \left\| {diag\left( {{{\bf{h}}_n}} \right){{\bf{g}}_n}} \right\|_2^2$ and ${Y_I} = \left\| {{{\bf{h}}_I}} \right\|_2^2$, ${Y_I}$ is also subjective to a Gamma distribution with the parameters of $\left( {K,{{\Omega _I}}} \right)$. From the derived process of Lemma \ref{CD-NOMA:lemma:gamma channel_CDF for far user}, the CDF ${F_{{Z_n}}}$ and PDF ${f_{{Y_I}}}$ are give by
\begin{align}\label{the CDF of Z_n}
{F_{{Z_n}}}\left( z \right)\approx  &  {\phi _n}\sum\limits_{p = 0}^{M - n} {{
 M - n \choose
 p
}\frac{{{{\left( { - 1} \right)}^p}}}{{n + p}}} \nonumber \\
  &\times {\left[ {\sum\limits_{u = 1}^U {{b_u}} \left( {1 - {e^{ - \frac{{z{c_u}}}{\eta }}}\sum\limits_{i = 0}^{K - 1} {\frac{1}{{i!}}{{\left( {\frac{{z{c_u}}}{\eta }} \right)}^i}} } \right)} \right]^{n + p}},
\end{align}
and
\begin{align}\label{the PDF of interference h_I}
{f_{{Y_I}}}\left( y \right) = \frac{{{y^{K - 1}}{e^{ - \frac{y}{{{\Omega _I}}}}}}}{{\left( {K - 1} \right){\rm{!}} \Omega _I^K}},
\end{align}
respectively. After some manipulations, \eqref{The CDF of the n-th user} can be rewritten as
\begin{align}\label{The CDF of the n-th user for the next step}
 {F_{{\gamma _n}}}\left( x \right) =& {\rm{Pr}}\left( {{Z_n} < \frac{{x\left( {\varpi \rho Y + 1} \right)}}{{\rho {a_n}}}} \right) \nonumber \\
  = & \int_0^\infty  {{f_Y}\left( y \right)} {F_{{Z_n}}}\left( {\frac{{x\left( {\varpi \rho y + 1} \right)}}{{\rho {a_n}}}} \right)dy .
\end{align}

Substituting \eqref{the CDF of Z_n} and \eqref{the PDF of interference h_I} into \eqref{The CDF of the n-th user for the next step}, we can easily obtain \eqref{CD-NOMA:the CDF of SINR expression for near user with ipSIC}, which completes the proof.

\appendices
\section*{Appendix~C: Proof of Theorem \ref{Theorem:CD-NOMA:the COP of near user for Case1 with ipSIC}} \label{Appendix:E}
\renewcommand{\theequation}{C.\arabic{equation}}
\setcounter{equation}{0}

Applying the assumptions in Appendix B of Lemma \ref{CD-NOMA:lemma:gamma channel_CDF for near user with SIC}, %\ref{Appendix:B}
we denote ${Z_n} = \left\| {diag\left( {{{\bf{h}}_n}} \right){{\bf{g}}_n}} \right\|_2^2 = \left\| {diag\left( {{{\bf{h}}_n}} \right){{\bf{g}}_m}} \right\|_2^2$ and ${Y_I} = \left\| {{{\bf{h}}_I}} \right\|_2^2$, respectively.
Substituting \eqref{The SINR expression at the n-th user to detect the m-th user} and \eqref{the SINR expression at n-th user to detect itself with SIC} into \eqref{CD-NOMA:the expression of COP for near user}, the COP of $P_{n1,CD}^{ipSIC}$ can be expressed as
\begin{align}\label{The derived expression of the n-th user with ipSIC for the case1}
P_{n1,CD}^{ipSIC} =& \underbrace {{\text{Pr}}\left( {\frac{{\rho {Z_n}{a_m}}}
{{\rho {Z_n}{a_n} + 1}} < {\varepsilon _m}} \right)}_{{J_1}} \hfill \nonumber \\
   & +  \underbrace {{\text{Pr}}\left( {\frac{{\rho {Z_n}{a_m}}}
{{\rho {Z_n}{a_n} + 1}} > {\varepsilon _m},\frac{{\rho {a_n}{Z_n}}}
{{\varpi \rho {Y_I} + 1}} < {\varepsilon _n}} \right)}_{{J_2}} \hfill ,
\end{align}
where ${J_1} = {F_{{Z_n}}}\left( \tau  \right)$, $\tau  = \frac{{{\varepsilon _m}}}{{\rho \left( {{a_m} - {\varepsilon _m}{a_n}} \right)}}$ with ${a_m} > {\varepsilon _m}{a_n}$ and $\varpi  = 1$.

After some mathematical manipulations, ${J_2}$ is calculated as
\begin{align}\label{The derived expression J_2 of the n-th user with ipSIC for the case1}
{J_2} = & {\text{Pr}}\left( {\tau  < {Z_n} < \vartheta {Y_I} + \beta } \right) \nonumber \\
    = & \underbrace {\int_0^\infty  {{f_{{Y_I}}}\left( y \right){F_{{Z_n}}}\left( {\vartheta y + \beta } \right)dy} }_{{J_3}} - {F_{{Z_n}}}\left( \tau  \right),
\end{align}
where $\beta {\text{ = }}\frac{{{\varepsilon _n}}}{{\rho {a_n}}}$ and $\vartheta {\text{ = }}\frac{{\varpi {\varepsilon _n}}}
{{{a_n}}}$.
%Substituting \eqref{The derived expression J_2 of the n-th user with ipSIC for the case1} into \eqref{The derived expression of the n-th user with ipSIC for the case1}, $P_{n1,CD}^{ipSIC}$ can be further formulated as follows:
%\begin{align}\label{The further derived expression of the n-th user with ipSIC for the case1}
%&P_{n1,CD}^{ipSIC} =  \int_0^\infty  {{f_{{Y_I}}}\left( y \right){F_{{Z_n}}}\left( {\vartheta y + \beta } \right)dy}.
%\end{align}
As can be seen from the above equation, it is pivotal to calculate the integral expression of \eqref{The derived expression J_2 of the n-th user with ipSIC for the case1}. Similar to the proving process of $Q_1$ in \eqref{The CDF of the n-th user for the next step}, based on \eqref{the CDF of Z_n} and \eqref{the PDF of interference h_I}, $J_3$ can be given by
\begin{align}\label{The derived expression J_3 of the n-th user with ipSIC for the case1}
&{J_3} \approx \frac{{{\phi _n}}}{{\left( {K - 1} \right) {\rm{!}}\Omega _I^K}}\sum\limits_{p = 0}^{M - n} {{
   {M - n}  \choose
   p
}\frac{{{{\left( { - 1} \right)}^p}}}{{n + p}}} \int_0^\infty  {{y^{K - 1}}{e^{ - \frac{y}{{{\Omega _I}}}}}}\nonumber  \\
&  \times {\left[ {\sum\limits_{u = 1}^U {{b_u}} \left( {1 - {e^{ - \frac{{{c_u}\left( {\vartheta y + \beta } \right)}}{\eta }}}\sum\limits_{i = 0}^{K - 1} {\frac{1}{{i!}}{{\left( {\frac{{\left( {\vartheta y + \beta } \right){c_u}}}{\eta }} \right)}^i}} } \right)} \right]^{n + p}}dy .
\end{align}

Substituting \eqref{The derived expression J_3 of the n-th user with ipSIC for the case1} and \eqref{The derived expression J_2 of the n-th user with ipSIC for the case1} into \eqref{The derived expression of the n-th user with ipSIC for the case1}, we can obtain \eqref{CD-NOMA:the COP of near user for Case1 with ipSIC} and complete the proof.

\appendices
\section*{Appendix~D: Proof of Theorem \ref{Theorem:CD-NOMA:the COP of near user with ipSIC for Case2}} \label{Appendix:F}
\renewcommand{\theequation}{D.\arabic{equation}}
\setcounter{equation}{0}

Similar to the variable substitutions in Appendix E of Theorem \ref{Theorem:CD-NOMA:the COP of near user for Case1 with ipSIC},
substituting \eqref{The SINR expression at the n-th user to detect the m-th user}, \eqref{the SINR expression at n-th user to detect itself with SIC} and \eqref{the SINR expression at n-th user to detect itself directly} into \eqref{OP n},
the CDF of $P_{CD,n2}^{ipSIC}$ can be expressed as
\begin{align}\label{The derived expression of the n-th user with ipSIC for the case2}
P_{n2,CD}^{ipSIC}=&  \underbrace {{\text{Pr}}\left( {\frac{{\rho {Z_n}{a_m}}}
{{\rho {Z_n}{a_n} + 1}} < {\varepsilon _m},\frac{{\rho {Z_n}{a_n}}}
{{\rho {Z_n}{a_m} + 1}} < {\varepsilon _n}} \right)}_{{J_1}} \hfill \nonumber\\
   &+ \underbrace {{\text{Pr}}\left( {\frac{{\rho {Z_n}{a_m}}}
{{\rho {Z_n}{a_n} + 1}} < {\varepsilon _m},\frac{{\rho {Z_n}{a_n}}}
{{\varpi \rho {Y_I} + 1}} < {\varepsilon _n}} \right)}_{{J_2}} \hfill  .
\end{align}

Followed by \eqref{The derived expression of the n-th user with ipSIC for the case2}, after some further manipulations, $J_1$ is formulated as
\begin{align}\label{The derived expression of the n-th user with ipSIC for the case2:D1}
{J_1} = {\text{Pr}}\left( {{Z_n} < \min \left( {\tau ,\upsilon } \right) \triangleq \zeta } \right) = {F_{{Z_n}}}\left( \zeta  \right),
\end{align}
where $\upsilon  = \frac{{{\varepsilon _n}}} {{\rho \left( {{a_n} - {\varepsilon _n}{a_m}} \right)}}$ with ${a_n} > {\varepsilon _n}{a_m}$. Substituting \eqref{the CDF of Z_n} into \eqref{The derived expression of the n-th user with ipSIC for the case2:D1}, $J_1$ is given by
\begin{align}\label{D1}
J_1  \approx & {\phi _n}\sum\limits_{p = 0}^{M - n} {{
 M - n \choose
 p
}\frac{{{{\left( { - 1} \right)}^p}}}{{n + p}}} \nonumber \\
  &\times {\left[ {\sum\limits_{u = 1}^U {{b_u}} \left( {1 - {e^{ - \frac{{\zeta{c_u}}}{\eta }}}\sum\limits_{i = 0}^{K - 1} {\frac{1}{{i!}}{{\left( {\frac{{\zeta{c_u}}}{\eta }} \right)}^i}} } \right)} \right]^{n + p}}.
\end{align}

Combining \eqref{D1}, \eqref{The derived expression J_2 of the n-th user with ipSIC for the case1} and \eqref{The derived expression J_3 of the n-th user with ipSIC for the case1}, \eqref{CD-NOMA:the COP of near user with ipSIC for Case2} can be obtained and
the proof is completed.

\bibliographystyle{IEEEtran}
\bibliography{mybib}

\end{document}